\documentclass[fleqn,usenatbib]{mnras}

\usepackage{newtxtext,newtxmath}
\usepackage[T1]{fontenc}

\DeclareRobustCommand{\VAN}[3]{#2}
\let\VANthebibliography\thebibliography
\def\thebibliography{\DeclareRobustCommand{\VAN}[3]{##3}\VANthebibliography}

\usepackage{graphicx}	% Including figure files
\usepackage{amsmath}	% Advanced maths commands
\usepackage{threeparttable}
\usepackage{booktabs}
\title[Strong \ion{H}{I} Absorption without galaxies]{\ion{H}{I} Absorbers as Beacons of Hidden Structure at $z \sim 3$: Multi-Component, Metal-Rich Absorption System near a Protocluster}

\author[K. Koretomo et al.]{
Kentaro Koretomo,$^{1}$\thanks{E-mail: krkn38@g.ecc.u-tokyo.ac.jp}
Nobunari Kashikawa,$^{1,2}$
Toru Misawa,$^{3}$
Jun Toshikawa,$^{4}$
Yoshihiro Takeda,$^{1}$
\newauthor
Hisakazu Uchiyama,$^{5,6}$
Junya Arita,$^{1}$
Shunta Shimizu,$^{1}$
Ryo Emori,$^{1}$
Kei Ito,$^{7, 8}$
Satoshi Kikuta,$^{9}$
\newauthor
Mariko Kubo,$^{4,10}$
Yoshiaki Ono,$^{11}$
Marcin Sawicki,$^{12}$
and Rhythm Shimakawa$^{13}$
\\
$^{1}$Department of Astronomy, School of Science, The University of Tokyo, 7-3-1 Hongo, Bunkyo, Tokyo 113-0033, Japan \\
$^{2}$Center for the Early Universe, The University of Tokyo, 7-3-1 Hongo, Bunkyo, Tokyo, 113-0033, Japan \\
$^{3}$Center for General Education, Shinshu University, 3-1-1 Asahi, Matsumoto, Nagano 390-8621, Japan \\
$^{4}$Astronomical Institute, Tohoku University, 6-3, Aramaki, Aoba, Sendai, Miyagi, 980-8578, Japan \\
$^{5}$Department of Advanced Sciences, Faculty of Science and Engineering, Hosei University, Koganei, Tokyo 184-8584, Japan \\
$^{6}$National Astronomical Observatory of Japan, Mitaka, Tokyo 181-8588, Japan \\
$^{7}$Cosmic Dawn Center (DAWN), Copenhagen, Denmark \\
$^{8}$DTU Space, Technical University of Denmark, Elektrovej 327, DK2800 Kgs. Lyngby, Denmark \\
$^{9}$Department of Regional Promotion, Nara Prefectural University, 10 Funahashicho, Nara, Nara, 630-8258, Japan \\
$^{10}$Department of Physics and Astronomy, School of Science, Kwansei Gakuin University, 1 Gakuen Uegahara, Sanda, Hyogo 669-1330, Japan \\
$^{11}$Institute for Cosmic Ray Research, The University of
Tokyo, 5-1-5 Kashiwanoha, Kashiwa, Chiba 277-8582, Japan \\
$^{12}$Institute for Computational Astrophysics and Department of Astronomy \& Physics, Saint Mary's University, 923 Robie Street, Halifax, NS B3H 3C3, Canada \\
$^{13}$Waseda Institute for Advanced Study (WIAS), Waseda University, 1-21-1, Nishi-Waseda, Shinjuku, Tokyo 169-0051, Japan
}

\date{Accepted XXX. Received YYY; in original form ZZZ}

\pubyear{\the\year{}}

\begin{document}
\label{firstpage}
\pagerange{\pageref{firstpage}--\pageref{lastpage}}
\maketitle

\begin{abstract}
\ion{H}{I} gas traces the large-scale structure and provides the primary fuel for star formation. High-$z$ protoclusters are ideal laboratories to study how \ion{H}{I} gas is accreted and consumed during the formation of the most massive structures in the Universe. However, much remains unknown about the distribution and physical state of their \ion{H}{I} gas. 
We examine a rare configuration in which a protocluster candidate is located in front of a quasar at $z=3.09$. Our spectroscopic campaign confirms a protocluster at $z = 3.079$: however, no corresponding strong \ion{H}{I} absorption is found in the background quasar spectrum.
Instead, we serendipitously discover a prominent \ion{H}{i} absorption feature at $z\sim 3.01$, offset by $\sim60$ cMpc from the centre of the protocluster. Spanning an exceptionally broad velocity range of $\sim 2000\,\mathrm{km\,s^{-1}}$ ($\sim 40\,\mathrm{cMpc}$), this absorption is decoupled from the confirmed member galaxies. Detailed kinematic modelling reveals this absorption comprises five distinct components rather than a single cloud. Moreover, one of these components exhibits a super-solar metallicity $(\mathrm{[O/H]} = +1.19^{+0.91}_{-0.78})$. 
We propose two physical scenarios for this unique system: (1) an additional, hidden massive protocluster along the line of sight, and/or (2) metal-rich outflows and metal-poor inflows driven by a single massive galaxy. The discovery highlights that while protoclusters are not universally associated with strong \ion{H}{I} absorption, targeting the strong \ion{H}{I} absorbers may serve as a beacon for uncovering massive, metal-rich protoclusters or complex gas kinematics in the early Universe.
\end{abstract}


%%%%%%%%%%%%%%%%%%%%%%%%%%%%%%%%%%%%%%%%%%%%%%%%%%

%%%%%%%%%%%%%%%%% BODY OF PAPER %%%%%%%%%%%%%%%%%%

\section{Introduction}
According to the hierarchical structure formation paradigm, small density fluctuations in the early Universe grow to form the immense cosmic web we observe today \citep[e.g.][]{Blumenthal+84, DEFW}. 
The earliest and most massive nodes of this network are called protoclusters, which are the ancestors of local massive galaxy clusters with $M_{\mathrm{halo}}^{z=0} > 10^{14} \, \mathrm{M_\odot}$ \citep{Overzier+16}. Because galaxy evolution is substantially accelerated in these highly overdense regions compared to the general field \citep[e.g.][]{Cen&Ostriker, Springel+06}, protoclusters offer unique astrophysical laboratories for understanding large-scale structure assembly and environmental drivers of galaxy evolution.
They are also important observational targets to elucidate the co-evolutionary pathways of galaxies and the supermassive black holes at their centres, because the environment may govern active galactic nucleus (AGN) triggering and subsequent star-formation quenching \citep[e.g.][]{Tozzi+22, Vito+24}. 
Their contribution to the cosmic star formation rate density is predicted to become increasingly dominant in the early Universe \citep{Chiang+17}. 
However, clusters of galaxies are rare in the local Universe, and discovering their progenitors at high redshifts is exceedingly difficult. 

To overcome these statistical limitations and truly understand rapid galaxy growth, tracing the stellar components alone is insufficient. It is imperative to map the neutral hydrogen (\ion{H}{I}) gas, as it exceeds the total stellar mass in the high-$z$ Universe \citep[e.g.][]{Peroux+21, Walter+20, Heintz+21} and traces the underlying dark matter distribution. Crucially, massive inflows of cold \ion{H}{I} gas from the intergalactic medium (IGM) act as the ultimate fuel reservoir driving intense galaxy formation within these structures \citep{Dekel+09}. One of the most powerful techniques to find protoclusters relies on tracking strong, coherent intergalactic Ly$\alpha$ absorption in the spectra of background quasars or galaxies \citep[e.g.,][]{Cai+17, Lee+18, Newman+25}. These massive absorption complexes arise from the overlapping of multiple Ly$\alpha$ absorption lines originating from the overdense diffuse IGM or the circumgalactic medium (CGM) of member galaxies.
This absorption-selected technique has gained significant attention following notable successes. For instance, \citet{Lee+18} established \ion{H}{I} tomography to reconstruct three-dimensional gas distributions using multiple background sightlines, successfully matching the highest absorption peak to a known overdensity of Ly$\alpha$ emitters (LAEs) at $z = 2.45$ \citep{Lee+16}. Similarly, \citet{Cai+16} detected massive overdensities at $z \sim 2.2$ by targeting regions with the highest Ly$\alpha$ optical depths in the eBOSS quasar sample \citep{Dawson+16}. Building on this context, \citet{Newman+22} recently detected a highly significant correlation between IGM Ly$\alpha$ transmission fluctuations and galaxy overdensities at $z \sim 2.5$. These successes underscore the critical importance of IGM tomography.
Moreover, recent \textit{JWST} observations have begun to map the \ion{H}{I} gas content of protoclusters at higher redshifts via Ly$\alpha$ absorption, as traced by background galaxies \citep{Heintz+26} and by the protocluster members themselves \citep[e.g.][]{Terp+26, Witten+26}.

Despite these observational milestones, both cosmological simulations and recent surveys have highlighted significant caveats in interpreting these absorption features. Using the EAGLE and Illustris simulations \citep{EAGLE, illustris}, \citet{Miller+19} predicted a weak correlation between mass overdensity and Ly$\alpha$ effective optical depth. They found that merely $\sim 55\%$ of robust \ion{H}{I} absorbers are physically associated with true protoclusters, while the remainder primarily trace extended filamentary structures. Furthermore, due to strong line-of-sight (LOS) dependencies, even the most massive protoclusters might evade detection in \ion{H}{I} absorption. Physical feedback further complicates this picture; \citet{Liang+21} found a deficit of cold IGM \ion{H}{I} in galaxy overdense regions, likely caused by intense radiation or preheating from member galaxies. Additionally, powerful AGN feedback can perturb the IGM on megaparsec scales, effectively increasing the \ion{H}{I} transparency within protoclusters \citep{Dong+24}. Observationally, this complexity is mirrored by findings such as a UV-dim protocluster candidate at $z \sim 2.5$ \citep{Newman+25}, which exhibits strong \ion{H}{I} absorption tracing the IGM but completely lacks a coincident overdensity of Lyman-break galaxies (LBGs) or LAEs. The existence of such discrepancies suggests that the traditional absorption-selected method might misidentify structures or miss certain massive overdensities entirely.

These conflicting observational and theoretical findings pose a fundamental question: are protoclusters ubiquitously associated with strong \ion{H}{I} gas? In addition, to what extent do absorption features in the spectra of background sources reliably map the spatial distribution of these structures? To answer these questions and overcome observational biases, we effectively reverse the conventional methodology in this study. Specifically, rather than searching for protoclusters starting from absorption features, we target a known, independently identified protocluster and investigate whether a significant amount of \ion{H}{I} absorption is actually observed in its vicinity. 
Recently, \citet[][]{Toshikawa+18,Toshikawa+24} conducted a systematic survey for protoclusters at $z = 3\text{--}5$ using data from the Hyper Suprime-Cam Subaru Strategic Programme \citep[HSC-SSP,][]{Aihara+18} and the CFHT Large Area $U$-band Deep Survey \citep[CLAUDS,][]{Sawicki+19}. Across a wide survey area of $\sim 25 \, \mathrm{deg^2}$, this study identified approximately 30 protocluster candidates per unit redshift by selecting regions that exhibit a significantly enhanced surface density of dropout galaxies. 
Among this extensive catalogue, we identified a uniquely compelling target: a protocluster candidate at $z \sim 3$ serendipitously located directly in front of a bright background SDSS quasar at $z=3.09$. 
Instead of using \ion{H}{I} absorption as a signpost to blindly search for a protocluster, we start with this photometrically selected galaxy overdensity and search the corresponding \ion{H}{I} absorption profile along the quasar sightline penetrating the protocluster. Such a spatial alignment is exceedingly rare. This is evident given that the comoving number densities of massive protoclusters and bright quasars at $z \sim 3$ are remarkably low: $\sim 10^{-5} \, \mathrm{cMpc^{-3}}$ \citep{Toshikawa+24} and $\sim 10^{-7} \, \mathrm{cMpc^{-3}}$ \citep{Ross+13}, respectively. Consequently, this fortuitous configuration provides a unique observational opportunity to investigate \ion{H}{I} gas distribution in a protocluster. 
Our results reveal a remarkably strong, multi-component absorption system at $z \sim 3.01$ that is kinematically decoupled from the spectroscopically identified protocluster.
In this paper, we present a detailed chemical and kinematic analysis of this rarely seen foreground absorber to unravel its physical origin.

This paper is organized as follows. In Section \ref{sec:data}, we describe the target selection and spectroscopic observations. Section \ref{sec:results} presents the analysis of the \ion{H}{I} and metal absorption lines, alongside the results of the photoionisation modelling. In Section \ref{sec:discussion}, we discuss the \ion{H}{I} absorption in the protocluster, investigate the physical origin of the strongest \ion{H}{I} absorption system, and evaluate potential scenarios. Finally, we summarise our main conclusions in Section \ref{sec:conclusions}. Throughout this paper, we adopt a standard $\Lambda$CDM cosmology with $\Omega_m = 0.3$, $\Omega_\Lambda = 0.7$, $H_0 = 70 \, \mathrm{km \, s^{-1} \, Mpc^{-1}}$, and $h=0.7$ and we use the AB magnitude system \citep{Oke&Gunn83}.

\section{Data}
\label{sec:data}
\subsection{Targets}
Our target is a massive overdensity of $U$-dropout galaxies with a significance of $4.15\sigma$, identified as ID11 by \citet{Toshikawa+24} in the DEEP 2-3 region of the HSC-SSP Deep layer. 
Please refer to \citet{Toshikawa+24} for information on how to select $U$-dropout galaxies and determine overdensities.  
This protocluster candidate is located along the line of sight toward the luminous background quasar SDSS J232729$-$004333 ($z = 3.09$). 
Notably, the SDSS DR17 \citep{SDSS_DR17} spectrum of the background quasar exhibits several strong \ion{H}{I} absorption features suggestive of a foreground structure. Furthermore, the photometric redshift ($z_{\mathrm{phot}}$) distribution of the $U$-dropout galaxies in this field, derived using \texttt{MIZUKI} \citep{Mizuki}, is skewed toward lower redshifts relative to the quasar. Specifically, the distribution peaks at $z \sim 3$, which is clearly lower than the quasar redshift ($z=3.09$) by $\Delta z \sim 0.1$.

\subsection{Observation}
Motivated by this rare alignment, which provides an ideal laboratory for directly probing the internal gas structure of a forming cluster, we conducted follow-up optical spectroscopy of this region using Subaru/FOCAS \citep{Kashikawa+02}. 
During this observational campaign, our targeting strategy was explicitly designed to map the \ion{H}{I} absorption across multiple sightlines, rather than relying solely on the single line of sight towards the background quasar. To achieve this, we allocated slits not only to the quasar and protocluster member candidates, but also to candidate background galaxies in the surrounding region. These targets were primarily selected from $U$-dropout galaxies using the colour-selection criteria outlined by \citet{Toshikawa+24}. HSC-SSP joint catalogue, which includes photometric redshifts determined by the Mizuki code \citep{Mizuki}. Specifically, candidates for both background galaxies and protocluster members were selected based on $2.8 < z_{\mathrm{phot}} < 3.6$ and $r < 24.6$, whereas additional protocluster member candidates were targeted using $2.8 < z_{\mathrm{phot}} < 3.1$ and $r < 26.38$.

The spectroscopic observations were carried out between 2023 September 3 and 5 (HST; proposal ID: S23B0007N) using the multi-object spectroscopy (MOS) mode of Subaru/FOCAS. We utilised the VPH520 grism with a 0.8 arcsec slit width, yielding a spectral resolution of $R \sim 1500$ over 4450--6050\,\AA. To accommodate the high density of $U$-dropout galaxies in the target field, we employed four MOS masks covering the same region, with some primary targets being observed in multiple masks. 
The total integration time for each object varies from 7,200 to 45,600 sec. 
As the background quasar serves as the primary probe for the foreground \ion{H}{I} absorption, it was included in all four MOS setups to maximise the total integration time (45,600 sec) and achieve the high signal-to-noise ratio (S/N) necessary for detailed absorption line analysis. 
For spectrophotometric calibration, we observed the hot subdwarf standard star BD+28~4211 ($\mathrm{R.A.} = 21^{\mathrm{h}}51^{\mathrm{m}}11^{\mathrm{s}}\!.05$, $\mathrm{Decl.} = +28^\circ51'50\farcs9$, J2000.0). The typical seeing size during the observation is $\sim 0\farcs5\text{--}0\farcs8$ based on the spatial profile of the standard star. Due to poor weather conditions on September 4, the total integration time obtained on this night was limited compared to the favourable nights of September 3 and 5.

A comprehensive summary of the observations is provided in Table \ref{tab:obs_overview}. 
Furthermore, we merged our data with an additional observational sample from Toshikawa et al. (2026, in prep.). The study investigates various properties of three protoclusters, including our target, selected by $U$-dropout overdensity in the HSC-SSP data.
To ensure consistency across these protoclusters, they restrict their final sample exclusively to galaxies satisfying the same $U$-dropout colour selection criteria. In contrast, our current analysis includes all spectroscopically confirmed objects, incorporating those selected via $z_{\mathrm{phot}}$. The spectra and the information of the sources not presented in Toshikawa et al. (2026, in prep.) are shown in Figure \ref{fig:all_spectra} and Table \ref{tab:obs_detail}. 
In total, we obtained spectra for $95$ $U$-dropout galaxies alongside the background quasar. Standard data reduction was performed using IRAF \citep{Doug+86, Doug+93}, which included bias subtraction, flat-fielding, wavelength calibration using a ThAr arc lamp, sky subtraction, 2D spectral combination, and flux calibration. Following the data reduction, we systematically inspected the spectra to identify emission and absorption features. We classified single emission lines exhibiting an asymmetric, redward-skewed profile as Ly$\alpha$ emission; given our spectral resolution of $R \sim 1500$, emission lines from low-redshift interlopers (e.g. [\ion{O}{II}]) would typically be resolved into doublets. For sources lacking prominent emission lines but exhibiting clear continuum, we searched for the Lyman break and interstellar medium (ISM) absorption features characteristic of LBGs. Finally, we spectroscopically confirmed 42 LAEs, 1 LBG, and the background quasar. 
Most of the remaining targets showed no signal, and no contamination from obvious low-$z$ emitters was found. 
The redshifts were determined from the peak wavelength of the Ly$\alpha$ emission for the LAEs, and from the wavelengths of the ISM absorption features and the Lyman break for the LBG not emitting Ly$\alpha$. 

\begin{table}
    \centering
    \caption{Overview of the observations.}
    \begin{tabular}{cccc} \hline \hline
        Date (HST) & Mask & $N_{\mathrm{slit}}$ & $t_{\mathrm{exp}}$ (min) \\ \hline
        3th September 2023 & a & 29 & 180 \\
                           & b & 31 & 120 \\
        4th September 2023 & c & 30 & 100 \\
        5th September 2023 & c & 30 & 200 \\
                           & d & 30 & 160 \\ \hline
    \end{tabular}
    \label{tab:obs_overview}
\end{table}

\begin{table}
    \centering
    \caption{Detailed information of the sources not presented in Toshikawa et al. (2026, in prep.)}
    \begin{threeparttable}
        \begin{tabular}{l c c c c}
            \hline \hline
            ID & RA & Dec & $z$ & $M_{\rm UV}$ \\
               & (J2000) & (J2000) & & (mag) \\
            \hline
            1 & 23:27:29.89 & -00:42:45.00 & 2.723 & -21.23 \\
            2 & 23:27:32.15 & -00:46:18.81 & 2.843 & -20.28 \\
            3 & 23:27:25.43 & -00:44:13.03 & 2.862 & -19.63 \\
            4 & 23:27:26.81 & -00:44:41.17 & 2.965 & -21.36 \\
            5\tnote{a} & 23:27:35.43 & -00:44:21.67 & 3.080 & -21.28 \\
            6 & 23:27:25.72 & -00:39:46.65 & 3.084 & -21.81 \\
            7 & 23:27:35.83 & -00:47:01.13 & 3.085 & -20.57 \\
            8 & 23:27:37.48 & -00:42:20.26 & 3.455 & -21.75 \\
            9 & 23:27:24.06 & -00:46:39.62 & 3.456 & -21.64 \\
            10 & 23:27:21.63 & -00:43:11.79 & 3.483 & -21.10 \\
            \hline
        \end{tabular}
        \begin{tablenotes}
            \footnotesize
            \item[a] LBG that lacks Ly$\alpha$ emission and exhibits only a Lyman break.
        \end{tablenotes}
    \end{threeparttable}
    \label{tab:obs_detail}
\end{table}

\section{Results}
\label{sec:results}
\subsection{Protocluster confirmation}
\begin{figure}
    \centering
    \includegraphics[width=\columnwidth]{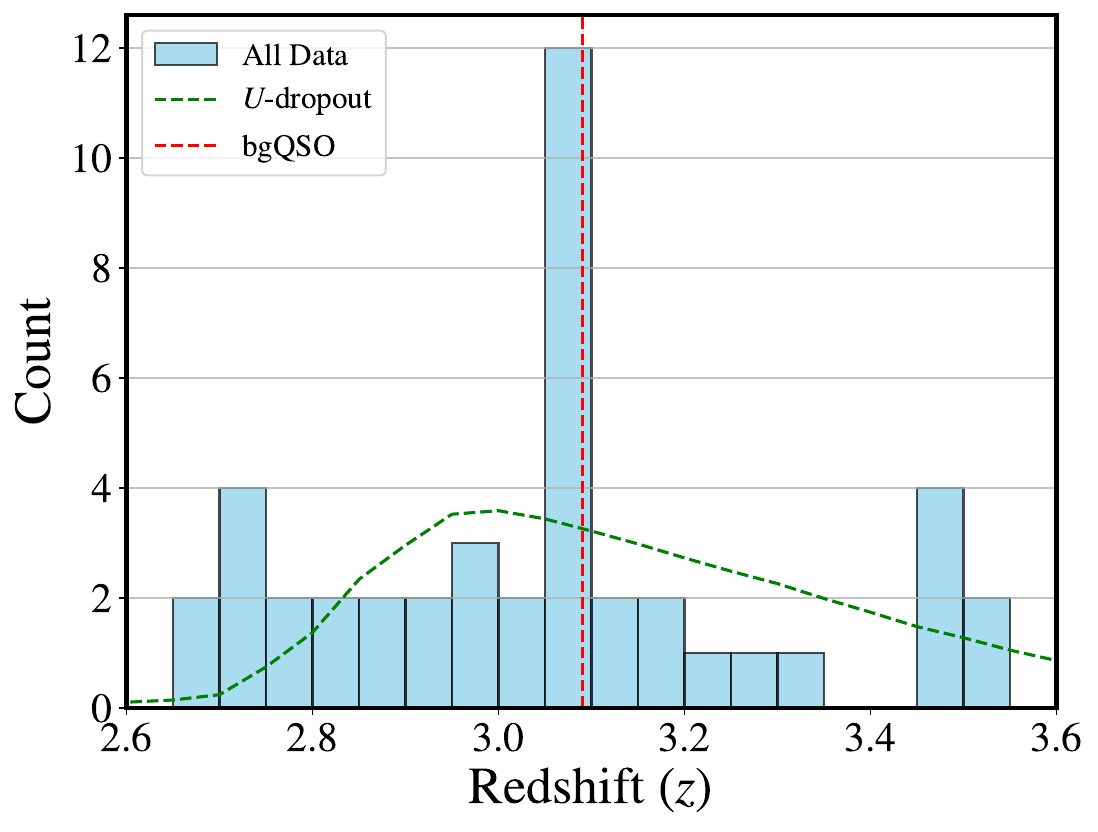}
    \caption{The redshift distribution of the spectroscopically confirmed galaxies and the quasar, whose redshift is shown in a red vertical dashed line. The green dashed line shows the expected spectroscopic redshift distribution of $U$-dropout galaxies \citep{Toshikawa+16} assuming a homogeneous distribution.}
    \label{fig:specz_distribution}
\end{figure}

\begin{figure}
    \centering
    \includegraphics[width=\columnwidth]{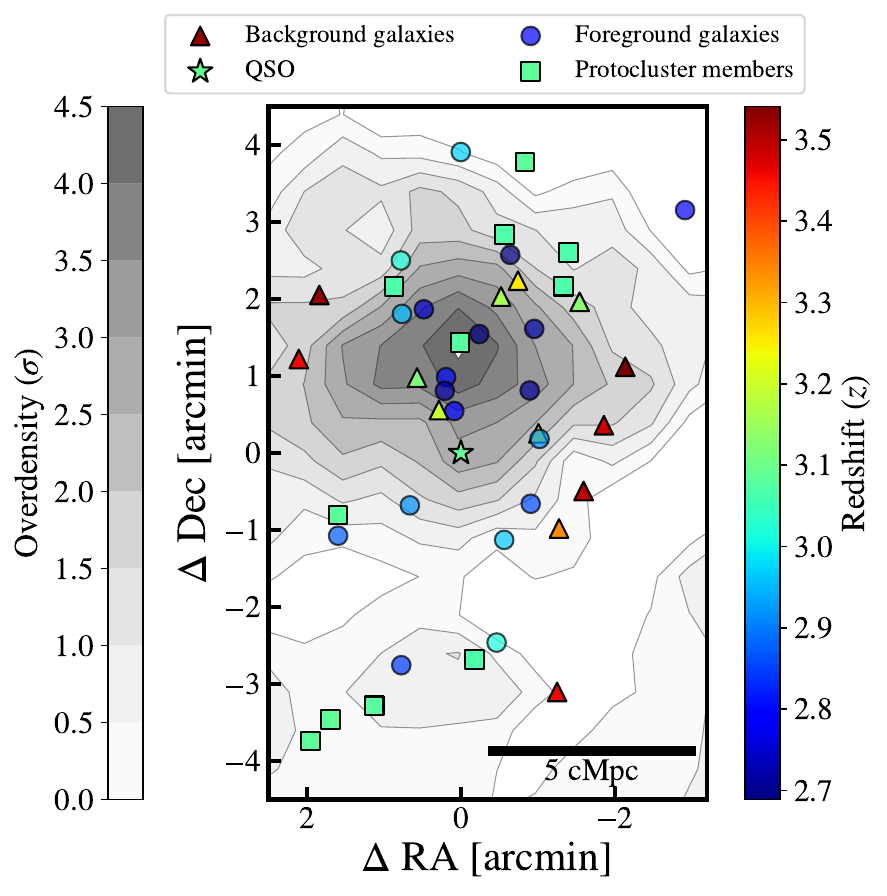}
    \includegraphics[width=\columnwidth]{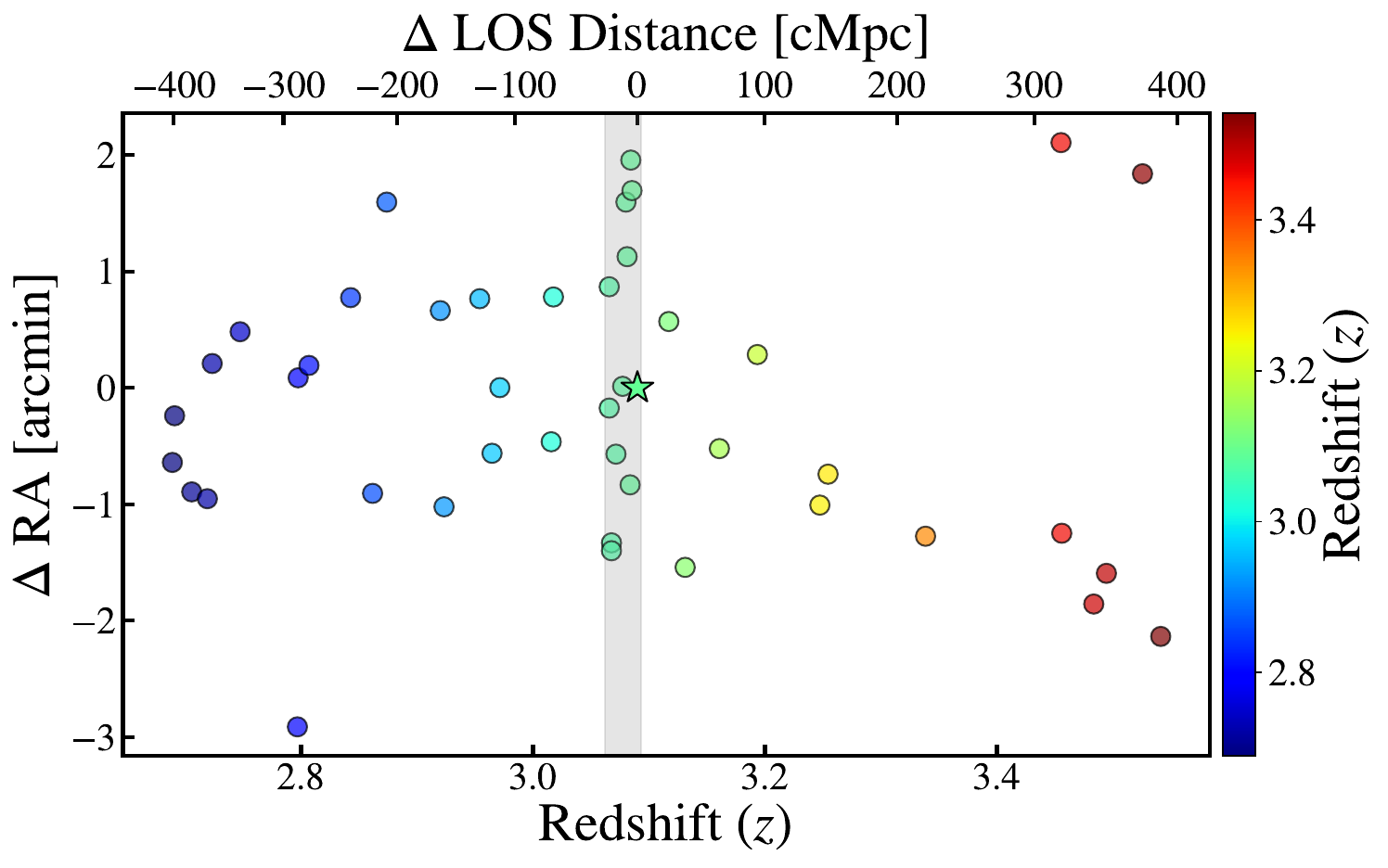}
    \caption{
    Upper panel: Sky distribution of the background galaxies (triangles), spectroscopically confirmed protocluster members (squares), and foreground galaxies (circles). The background quasar is marked by a star at the coordinate origin. All symbols are colour-coded by redshift. The underlying grey contours display the $U$-dropout galaxy overdensity, spanning from $0\sigma$ to $4.5\sigma$ in $0.5\sigma$ intervals. The black scale bar denotes a comoving length of $5\,\mathrm{cMpc}$ at $z=3.09$. 
    Lower panel: RA versus redshift for the same sources, using the same symbols as in the upper panel. The top horizontal axis indicates the relative LOS distance from the quasar in comoving units. The grey-shaded region highlights the redshift range of the protocluster, slightly broadened for visualization.
    }
    \label{fig:sky_dist_relative}
\end{figure}

Figure \ref{fig:specz_distribution} presents the spectroscopic redshift distribution of our sample, revealing a prominent overdensity ($N=12$) in the range $z = 3.05\text{--}3.10$ including the quasar. 
Assuming a homogeneous distribution of field $U$-dropout galaxies \citep{Toshikawa+16}, this structure exhibits a significant overdensity of $\delta = 2.49$, where $\delta \equiv (N_{\mathrm{obs}} - N_{\mathrm{exp}})/N_{\mathrm{exp}}$. 
Specifically, $N_{\mathrm{exp}}$ is the expected number of detections from spectroscopic observations of $U$-dropout galaxies, derived by integrating a homogeneous distribution over the redshift range of $z=3.05\text{--}3.10$. $N_{\mathrm{obs}}$ is the number of galaxies observed in the same redshift range. Before this integration, the distribution is normalised so that its total integrated area matches our total number of spectroscopically confirmed galaxies times $\Delta z = 0.05$. This calculation yields $N_{\mathrm{exp}} = 3.44$. Based on the expected number, the probability of finding 12 or more galaxies at this $\Delta z$ by chance is $2.5 \times 10^{-4}$ when assuming a homogeneous distribution.
We conclude that we have spectroscopically confirmed a protocluster at $z=3.079$, defined as the mean of the minimum and maximum member redshifts, with a redshift span of $\Delta z=0.028$ ($\sim 26 \, \mathrm{cMpc}$ along LOS). Figure \ref{fig:sky_dist_relative} shows the sky distribution of the spectroscopically confirmed sample, as well as the redshift versus RA distribution. 
The physical position of the quasar is at the deepest edge of the protocluster.
Therefore, the protocluster lies entirely in the foreground relative to the quasar, which validates our methodology to analyse any associated \ion{H}{i} absorption lines.
Toshikawa et al. (2026, in prep.) derive physical properties such as the absolute UV magnitude ($M_{\mathrm{UV}}$), stellar mass ($M_*$), and dust reddening ($E(B-V)$) for member galaxies based on SED fitting of the multi-band photometric data.
However, no significant correlation between these physical parameters and the spatial structure of the protocluster is found.

\subsection{Ly\texorpdfstring{$\alpha$}{a} transmission vs galaxy distribution}
\label{subsec:transmission}
\begin{figure}
    \centering
    \includegraphics[width=\columnwidth]{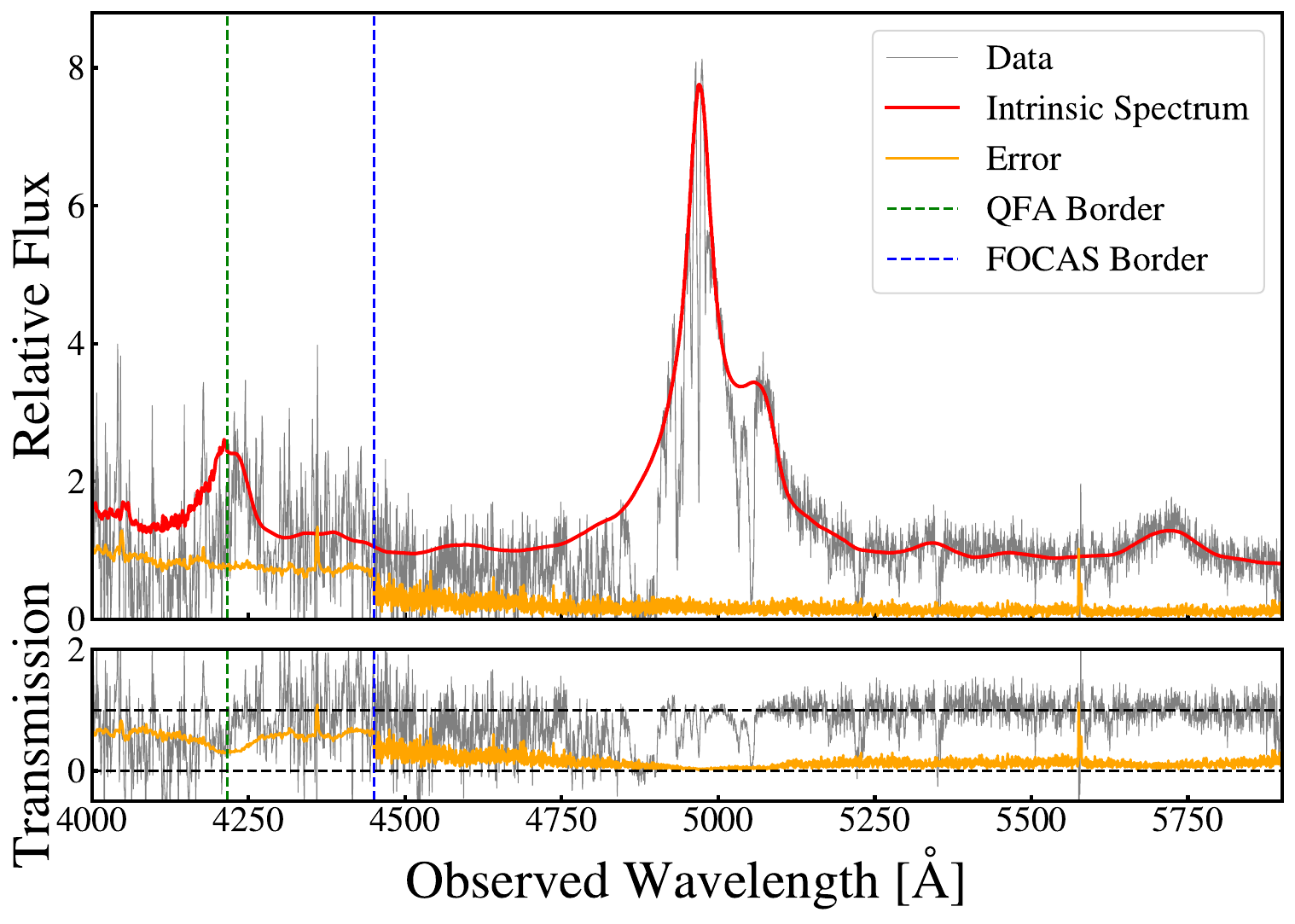}
    \caption{Intrinsic spectrum of the quasar estimated by using QFA \citep{Sun+23} and \textsc{SpenderQ} \citep{SpenderQ}. Upper panel: The observed spectrum and the estimated intrinsic spectrum are shown as grey and red lines, respectively. The error spectrum is plotted as an orange line. 
    The \textsc{SpenderQ} model is used for wavelengths shorter than the green vertical dashed line, while the QFA model is used for longer wavelengths.
    SDSS data are used for wavelengths shorter than the blue vertical dashed line, and FOCAS data for longer wavelengths.
    Lower panel: The derived transmission spectrum and its associated error are represented by grey and orange lines, respectively.}
    \label{fig:QFA_fit}
\end{figure}

\begin{figure}
    \centering
    \includegraphics[width=\columnwidth]{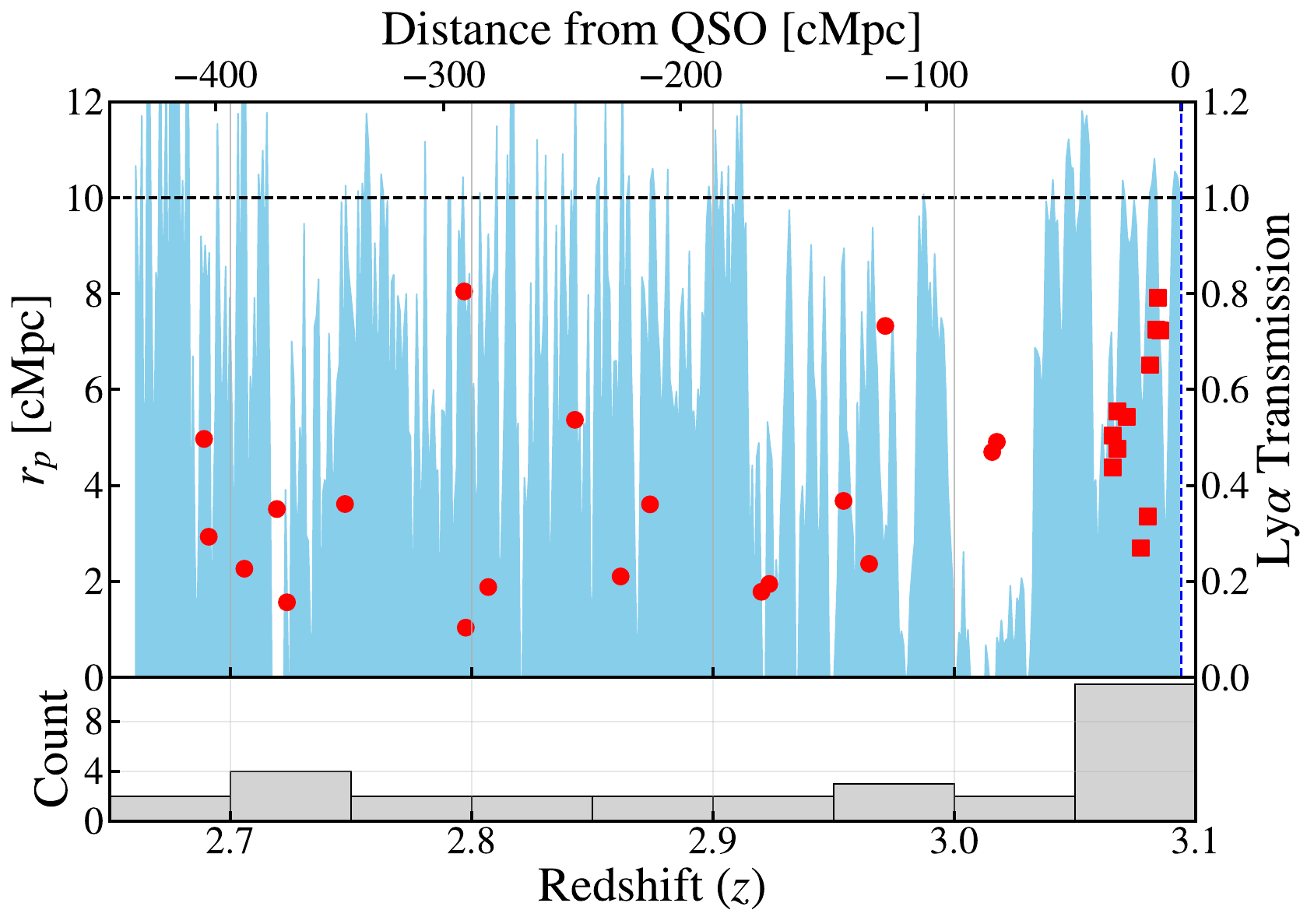}
    \caption{Ly$\alpha$ transmission of the quasar (sky-blue shaded region) and comoving impact parameters ($r_{\mathrm{p}}$) of spectroscopically confirmed galaxies from the quasar sightline. Protocluster members are indicated by filled red squares, while non-member galaxies are shown as filled red circles. Note that the $x$- and $y$-axes have different comoving distance scales. The lower panel shows a redshift distribution of the spectroscopically confirmed galaxies.
    }
    \label{fig:transmission_galaxy}
\end{figure}

To investigate the correspondences between the \ion{H}{I} gas and the member galaxies, we derive the Ly$\alpha$ transmission along the background quasar sightline. The intrinsic quasar spectrum, free from IGM absorption, is inferred using Quasar Factor Analysis \citep[QFA\footnote{\url{https://github.com/ZechangSun/QFA}};][]{Sun+23}. QFA combines factor analysis with physical priors of the IGM to predict the intrinsic quasar continuum within the Ly$\alpha$ forest to an accuracy of $\sim 2\%$. When applying this method to our quasar spectrum, we adopt the redshift of $z = 3.094$ derived from the SDSS catalogue, as it is difficult to reliably determine the systemic redshift directly from Ly$\alpha$ emission. During this fitting process, we also incorporate wavelength masks to exclude the strongest \ion{H}{i} absorption and intervening metal absorption lines unassociated with the quasar. 
The resulting fit is shown in Figure~\ref{fig:QFA_fit}, and it can be seen that the fit is overall quite good.
We note that the fit is relatively poor near the absorption blueward of the \ion{N}{v} emission line (5020--5060\AA \, in observed wavelength) because this region is masked due to contamination unrelated to the quasar itself. However, this localised deviation has a negligible effect on our subsequent analysis. 
Strong absorption can apparently be seen at the observed wavelength of $\sim4880$\,\AA. 

Figure \ref{fig:transmission_galaxy} compares the derived transmission profile with the spatial distribution of spectroscopically confirmed galaxies along the quasar sightline. 
We highlight two key findings from this comparison: first, the \ion{H}{I} absorption at the redshift of the protocluster at $z\sim3.08$ is relatively weak; second, the strongest \ion{H}{I} absorption system at $z\sim 3.01$, which extends over 40 cMpc along the sightline, does not coincide with this protocluster.
Although two galaxies are identified at the redshift of this strong absorption feature, their impact parameters from the quasar sightline exceed $1 \, \mathrm{pMpc}$. According to the empirical relation between the impact parameter and Ly$\alpha$ equivalent width established by \citet{Rakic+12}, a distance exceeding $1 \, \mathrm{pMpc}$ is too large to account for such a prominent absorption system with a rest-frame equivalent width of $EW_0 > 10~\text{\AA}$.

\subsection{The \texorpdfstring{\ion{H}{I}}{HI} gas in the protocluster}
\label{subsec:HIgas_PC}
\begin{figure}
    \centering
    \includegraphics[width=\columnwidth]{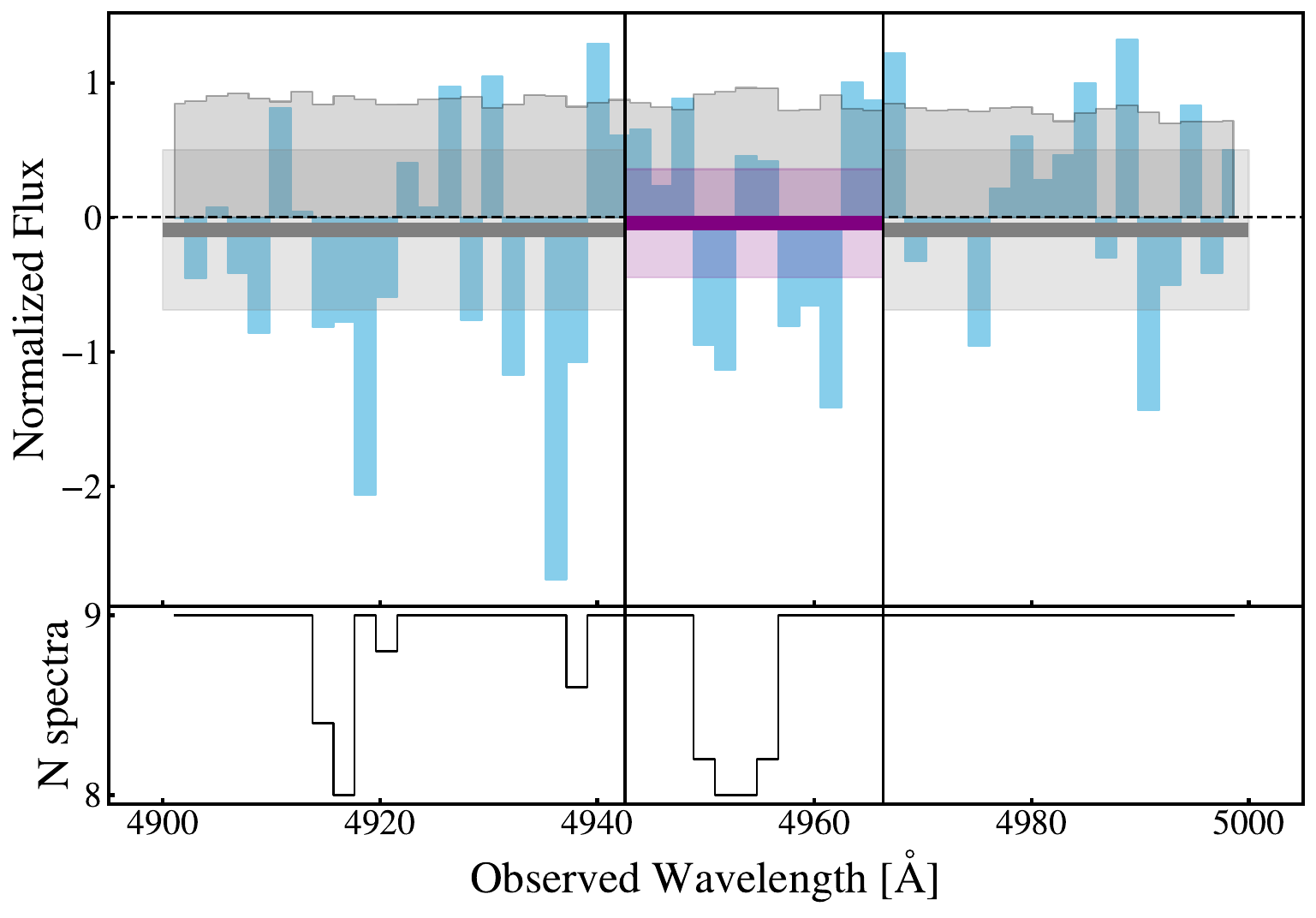}
    \caption{Median-stacked spectrum of the background galaxies ($N=9$). The vertical black lines indicate the redshift range corresponding to the protocluster. The grey shaded region represents the $1\sigma$ uncertainty level. The thick horizontal lines denote the mean flux levels within each respective wavelength region; the protocluster region is shown in purple, and outside regions are shown in grey.
    For visualization purposes, the spectrum is binned by 5 pixels.
    The lower panel shows the histogram of the number of spectra used for stacking at each wavelength, which has also been rebinned into the same 5-pixel bins.}
    \label{fig:stacked_PC}
\end{figure}

As shown in Figure \ref{fig:transmission_galaxy}, there are some weak \ion{H}{i} absorption features near the protocluster redshift ($z \sim 3.06$, compared to $z \sim 3.08$ for the protocluster centre). However, Figure \ref{fig:transmission_galaxy} indicates that the confirmed protocluster members have large impact parameters, with the smallest being $2.69\,\mathrm{cMpc}$. Although undetected faint member galaxies, which may be located near the quasar sightline, can cause this absorption, we cannot currently distinguish whether these features originate from faint galaxies or the protocluster \ion{H}{i} gas. 
To further investigate the overall \ion{H}{i} absorption associated with the protocluster, we stack all available background source spectra to maximise the S/N. 
The individual spectra are normalised by their rest-frame flux at $1280$\,\AA. Four galaxies are excluded from this procedure: two due to insufficient wavelength coverage and two due to negative normalisation factors, leaving a final sample of 9 galaxies at $z = 3.117\text{--}3.525$ for the stacking analysis. Before stacking, we exclude pixels with errors greater than three times the median error of each spectrum. To minimise the influence of outliers, we adopt median stacking. To estimate the uncertainty, we perform bootstrap resampling. In each realisation, we randomly select the same number of spectra as in the original sample, with replacement. The selected spectra are then median-stacked, and the mean flux levels are calculated separately for the protocluster and non-protocluster wavelength regions. 
Figure \ref{fig:stacked_PC} shows the result of the stacking analysis, which reveals no significant \ion{H}{I} absorption at the redshift of the protocluster. The horizontal lines indicate the mean flux levels measured from the original median-stacked spectrum. 
We then compute the difference between these two mean flux levels. This procedure is repeated 10,000 times to obtain the distribution of the flux difference. We adopt the median of this distribution as the flux difference and its standard deviation as the 1$\sigma$ uncertainty. As a result, we obtain a flux difference (non-protocluster minus protocluster) of $0.018 \pm 0.37$. Therefore, we do not find any significant difference between the non-protocluster and protocluster regions.

This non-detection might be due to the large impact parameters between the background sightlines and the confirmed protocluster members (with a mean impact parameter of $6.13$~cMpc)
, compounded by the low S/N of the individual spectra. Consequently, we are unable to place strong constraints on the global \ion{H}{I} gas distribution within this structure. 
In Section~\ref{subsec:absence_HI}, we discuss potential reasons for the absence of \ion{H}{I} absorption in this protocluster. 

\subsection{The strongest \texorpdfstring{\ion{H}{I}}{HI} absorption}
\label{subsec:strongest_HI}
As noted in Section \ref{subsec:transmission}, the strongest \ion{H}{I} absorption feature in the quasar spectrum lacks a detected galaxy counterpart. Initially, we hypothesise that this feature originates from a single Damped Lyman-$\alpha$ (DLA) system, typically defined by an \ion{H}{I} column density of $\log N_{\text{H\,\textsc{i}}} \, [\mathrm{cm^{-2}}] \ge 20.3$ \citep{Wolfe+05}. However, three lines of evidence suggest that this system is unlikely to be a canonical single DLA. 

\begin{figure}
    \centering
    \includegraphics[width=\columnwidth]{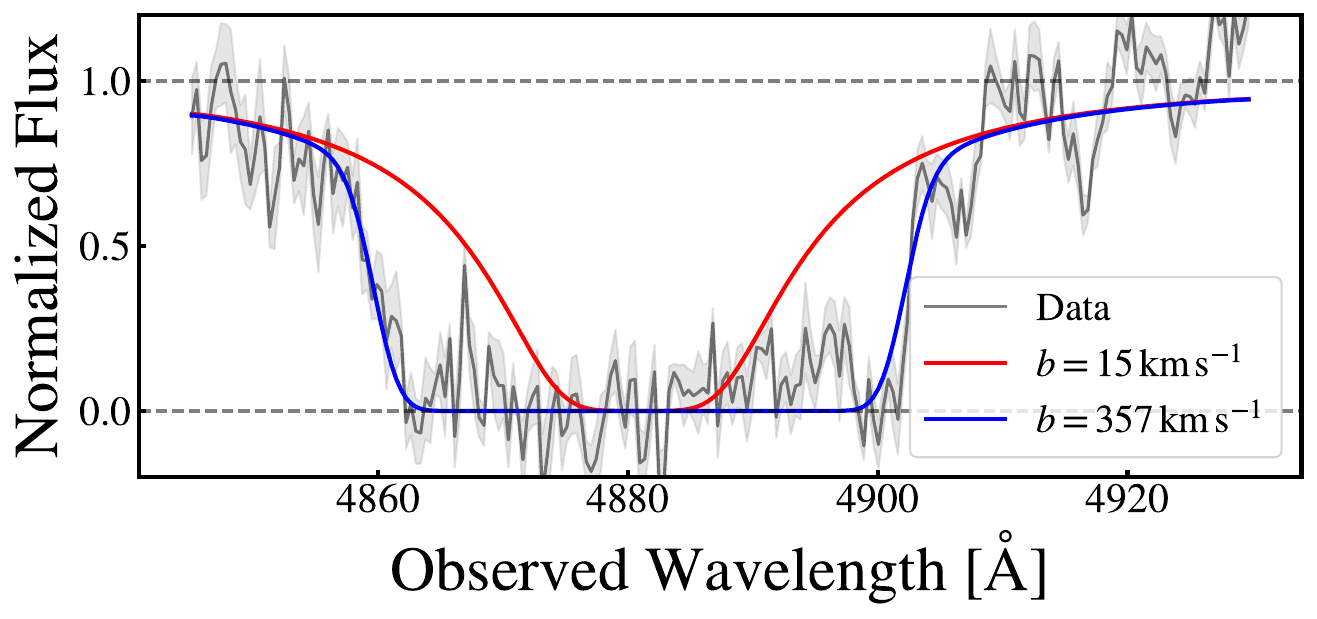}
    \caption{Comparison between the observed data (grey line) and the model spectra. We present two cases: one assuming a typical Doppler parameter of $b = 15\,\mathrm{km\,s^{-1}}$ (red), and another using the unconstrained best-fit value of $b = 357\,\mathrm{km\,s^{-1}}$ (blue). Both models adopt a fixed column density of $\log N_{\mathrm{H\,\textsc{i}}} [\mathrm{cm^{-2}}] = 20.28$. Note that both this column density and the large Doppler parameter ($b = 357\,\mathrm{km\,s^{-1}}$) are simultaneously derived from an MCMC analysis where $z$, $N_{\mathrm{H\,\textsc{i}}}$, and $b$ are left as free parameters.}
    \label{fig:DLA_fit}
\end{figure}

First, the absorption profile lacks the extended Lorentzian damping wings characteristic of a DLA, exhibiting instead the steep, abrupt edges more indicative of a blended, saturated IGM absorption complex. Figure \ref{fig:DLA_fit} compares the observational data with two distinct Voigt profile models. The first is an unconstrained best-fit model, while the second represents a more realistic case, assuming a typical Doppler parameter but retaining the column density derived from the unconstrained analysis.
Attempting to fit this strong absorption with a single Voigt profile without parameter constraints results in an unphysically larger Doppler parameter ($b \sim 360 \, \mathrm{km \, s^{-1}}$). If dominated by thermal motion, such a broad $b$-value implies a gas temperature exceeding $10^6 \, \mathrm{K}$, at which point hydrogen would be almost entirely collisionally ionised. Second, we find a substantial kinematic offset of $\sim 410 \, \mathrm{km \, s^{-1}}$ between the primary \ion{H}{I} absorption centre ($z=3.015$) and the associated metal absorption lines (\ion{O}{I} $\lambda$1302; \ion{Si}{II} $\lambda\lambda$1260, 1304; \ion{Si}{IV} $\lambda\lambda$1393, 1402; see Section \ref{subsubsec:metal_line_identification}) at $z=3.009$. Third, adopting the solar abundance scale of \citet{Asplund+21}, the metallicity derived from these column densities ($\mathrm{[O/H]} \approx \log(N_{\mathrm{O\,I}} / N_{\mathrm{H\,I}}) - \log(\mathrm{O/H})_\odot \approx +0.3$) is high for a DLA, placing this system near the upper envelope of the typical high-$z$ DLA population \citep{Quiret+16}. We conclude that this entire strong absorption system is unlikely to originate from a single DLA.

\subsubsection{Multi-component hypothesis}
Next, we assess the hypothesis that the observed \ion{H}{I} absorption feature is a blend of multiple discrete clouds, assuming the CGMs. Because the Ly$\alpha$ absorption is strongly saturated (reaching zero transmission), incorporating the corresponding Ly$\beta$ transmission profile is essential to robustly constrain the column densities and Doppler parameters in a simultaneous Voigt profile fit. 
Since the FOCAS data do not include the wavelength range of 4100--4150\,\AA \, corresponding to Ly$\beta$, we use the SDSS spectrum instead.
In addition, the predictive wavelength range of QFA, which is used to estimate the intrinsic quasar continuum, does not extend sufficiently blueward to cover the Ly$\beta$ transition, we incorporate an alternative machine-learning framework, \textsc{SpenderQ}\footnote{\url{https://github.com/galactic-ai/SpenderQ}} \citep{SpenderQ}. \textsc{SpenderQ} has been demonstrated to reconstruct the intrinsic continuum blueward of Ly$\alpha$ with an accuracy better than $5\%$. 
While \textsc{SpenderQ} effectively predicts the continuum redward of the Ly$\alpha$ emission, it fails to provide a robust estimate near the quasar emission lines (Ly$\alpha$, \ion{N}{V}, and \ion{Si}{IV}) for our specific dataset. 
In contrast, QFA provides a superior overall fit to the spectrum. 
Therefore, to obtain an optimal, full-wavelength intrinsic spectrum, we construct a composite model by primarily retaining the QFA reconstruction and utilising the \textsc{SpenderQ} reconstruction only outside the QFA estimation range, such as in the Ly$\beta$ region. Although there is an offset between QFA and \textsc{SpenderQ}, which exhibit a mean discrepancy of 13\% in the Ly$\alpha$ forest continuum region (4350--4700~\AA), this stems from methodological differences between the tools and does not affect our final results. 
Using this composite intrinsic spectrum to normalise the observed spectrum, we then proceed with the multi-component fitting. 

We perform the fitting procedure using the Python package \texttt{lmfit} \citep{lmfit}, managing the atomic data via \texttt{linetools} \citep{linetools}. To account for instrumental broadening, the models are convolved with a line spread function (LSF) corresponding to a spectral resolution of $R \sim 1500$ for the FOCAS data and $R \sim 1650$ for the SDSS data \citep{Smee+13}. 
We estimate the posterior distributions using a Markov Chain Monte Carlo (MCMC) approach implemented in \texttt{emcee} \citep{Foreman-Mackey+13}, running 5000 steps following a burn-in phase of 500 steps. We systematically evaluate models with multiple components to determine the optimal number of discrete absorbing clouds. 
Given the presence of multiple associated metal absorption lines firmly detected at $z=3.009$ (see Section~\ref{subsubsec:metal_line_identification}), we fix the redshift of the primary metal-bearing component at $z=3.009$ and introduce a variable number of additional kinematic components. 
The column density ($N$), Doppler parameter ($b$), and redshift ($z$) of each additional component are treated as free parameters. 
We first apply a 1D Gaussian filter with $\sigma = 1.5\,\mathrm{pixels}$ to the residual spectrum (data minus model) to mitigate the impact of local noise fluctuations. Additional components are then sequentially introduced at the redshifts corresponding to the most negative values in the smoothed residual, restricting the search to regions exhibiting saturated absorption. The initial parameters for these new components are derived from the previous fit. 

Furthermore, the range of the Doppler parameter $b$ is physically restricted. According to \citet{Tumlinson+17}, the non-thermal component $b_{\mathrm{nt}}$ of CGM is $b_{\mathrm{nt}} < 20 \, \mathrm{km \, s^{-1}}$ for low-ionization species and $b_{\mathrm{nt}} \sim 50\text{--}75 \, \mathrm{km \, s^{-1}}$ for high-ionization species. In addition to $b_{\mathrm{nt}}$, the gas temperature $T$ also governs the total $b$. \citet{Tumlinson+17} state that $T \sim 10^4 \text{--} 10^5 \, \mathrm{K}$ for low ions and $T \sim 10^{5.7} \, \mathrm{K}$ for high ions. Consequently, combining these extreme values yields a theoretical lower limit of $b=13 \, \mathrm{km \, s^{-1}}$ (assuming $b_{\mathrm{nt}} = 0 \, \mathrm{km \, s^{-1}}$ and $T=10^4 \, \mathrm{K}$) and an upper limit of $b=118 \, \mathrm{km \, s^{-1}}$ (assuming $b_{\mathrm{nt}}=75 \, \mathrm{km \, s^{-1}}$ and $T \sim 10^{5.7} \, \mathrm{K}$). 
Based on this, we constrain the prior range to $10 < b < 120 \, \mathrm{km \, s^{-1}}$ for the metal-associated component at $z=3.009$. For all other components, we impose a range of $10 < b < 50 \, \mathrm{km \, s^{-1}}$, which is consistent with the theoretical upper limit for low ions ($b=45 \, \mathrm{km \, s^{-1}}$ given $T=10^5 \, \mathrm{K}$ and $b_{\mathrm{nt}} = 20 \, \mathrm{km \, s^{-1}}$). Because our primary aim in the subsequent iterations is to successfully inherit $z$ and the column density $N$, we initialize $b$ at a typical value of $b=30 \, \mathrm{km \, s^{-1}}$. Finally, the initial value for $z$ is inherited from the previous fit, and we allow it to vary within a window of $\Delta z = 0.002$ for all components except the fixed one at $z=3.009$. Aside from the aforementioned boundary restrictions, we do not impose any informative priors; instead, we adopt flat priors for all parameters within their respective ranges. The fitting utilizes 276 data points. For a model with $n$ components, the number of free parameters is $2 + 3(n - 1)$, which yields $277 - 3n$ degrees of freedom.

\begin{table}
    \centering
    \caption{Statistical evaluation of the multi-component Voigt profile fitting.}
    \begin{tabular}{c c c r} \hline \hline
         Number of components & Reduced $\chi^2$ & BIC & $\Delta$BIC \\ \hline
         1 & 10.8 & 666 & 0        \\
         2 & 4.62 & 445 & $-221$   \\
         3 & 3.99 & 419 & $-26.31$ \\
         4 & 3.89 & 426 & $+6.93$  \\
         5 & 3.36 & 399 & $-27.09$ \\ 
         6 & 3.26 & 404 & $+5.12$  \\ 
         7 & 3.31 & 422 & $+17.98$ \\ \hline
    \end{tabular}
    \label{tab:evaluation}
\end{table}

\begin{figure}
    \centering
    \includegraphics[width=\columnwidth]{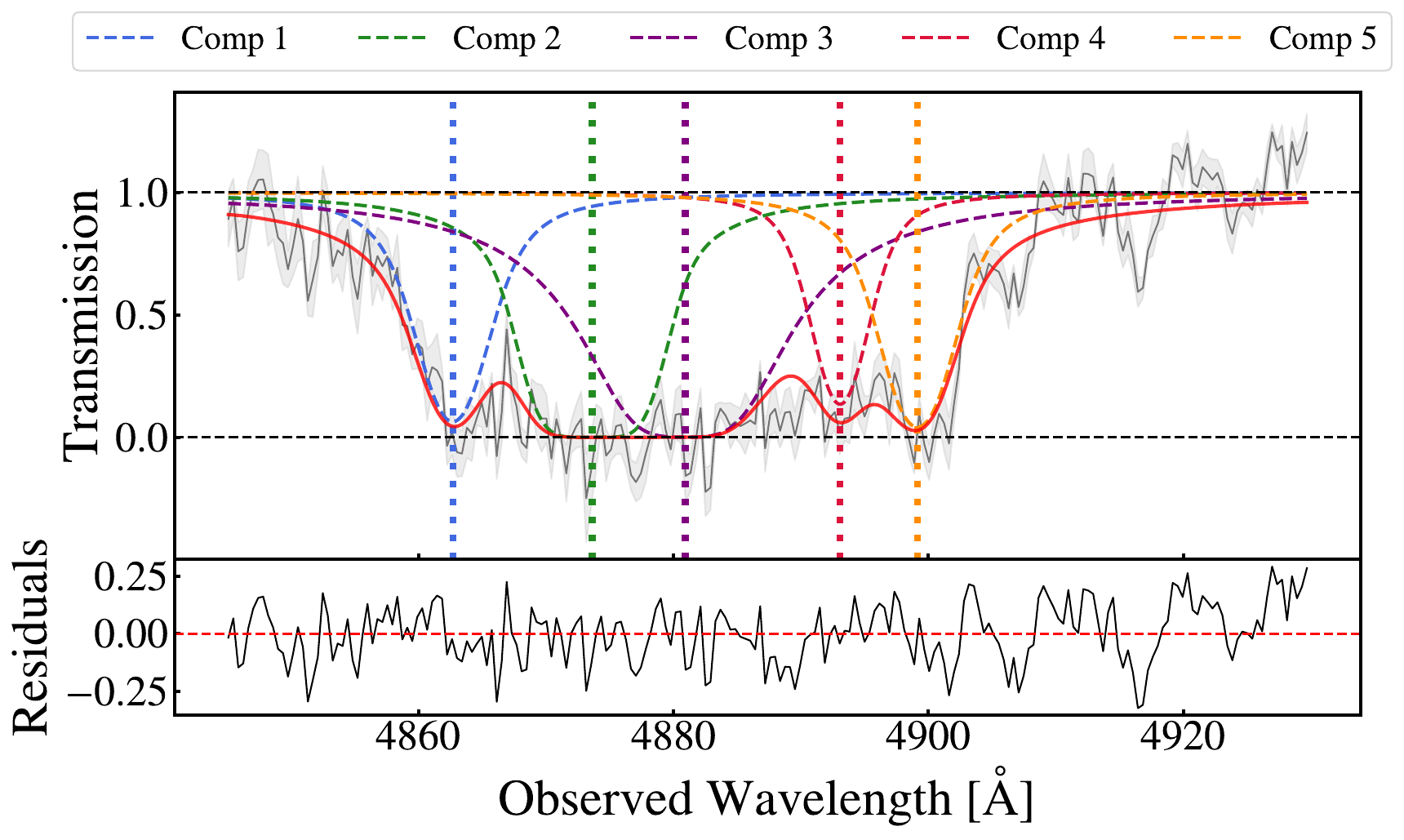}
    \includegraphics[width=\columnwidth]{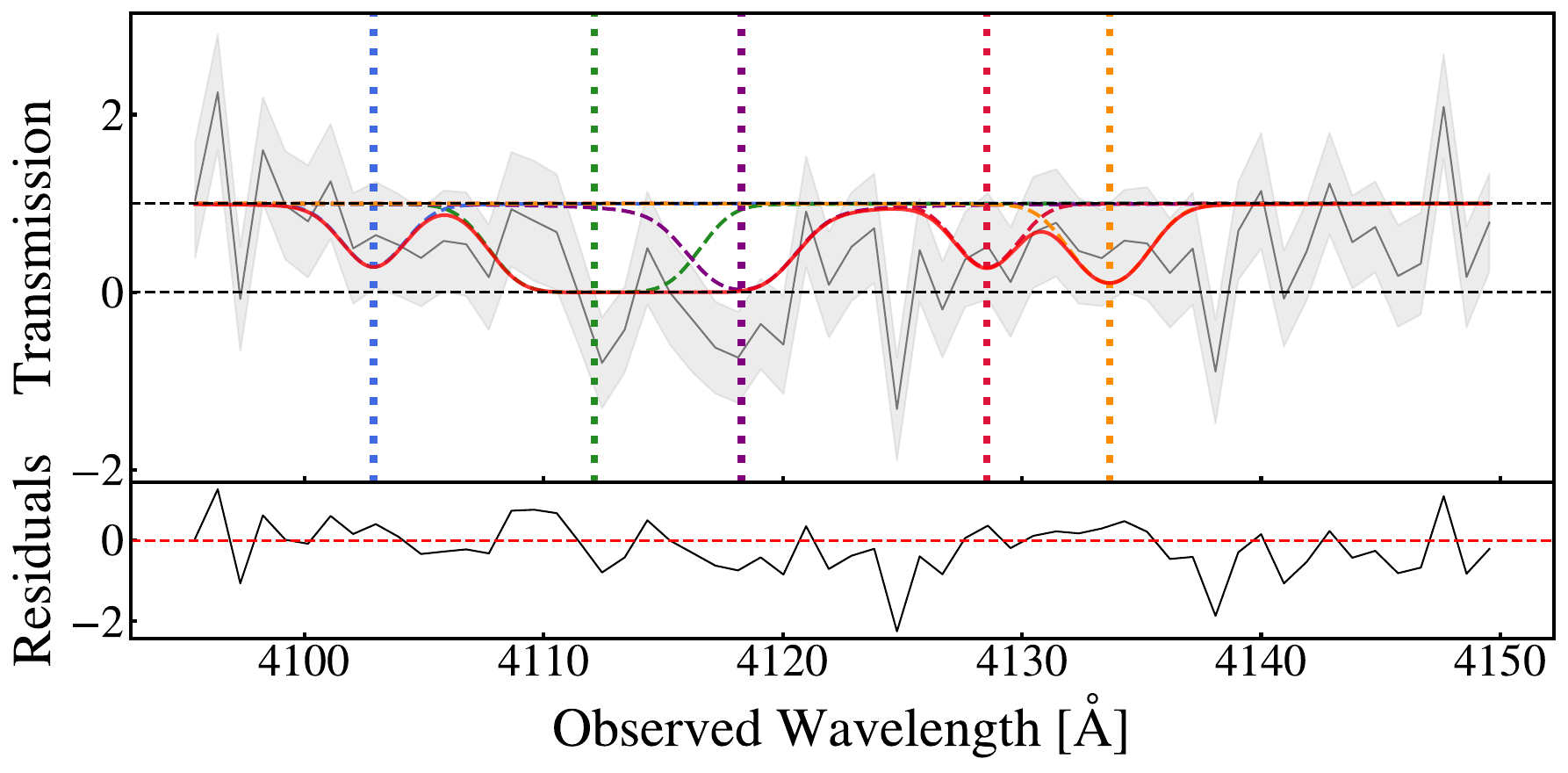}
    \caption{The best-fit five-component Voigt profile model. 
    Transmission profiles and the corresponding best-fit models for the Ly$\alpha$ (upper panel) and Ly$\beta$ (lower panel) transitions.
    The grey line and the shaded region indicate the observed transmission with its $1\sigma$ uncertainties. 
    Each coloured dotted line corresponds to a specific component, whose kinematic parameters are listed in Table \ref{tab:5component_fit_parameter}, and their superimposition is shown as a red line.
    The sub-panels below each transmission curve display the fitting residuals. Note that the redshift of the primary metal-bearing component at $z=3.009$ is fixed during the fitting procedure.}
    \label{fig:5component_fit_picture}
\end{figure}

Table~\ref{tab:evaluation} summarises the goodness-of-fit for each model based on the reduced $\chi^2$ statistic, the Bayesian Information Criterion (BIC) \citep{BIC}, and the $\Delta$BIC compared to the previous fit. Although we extend our fitting to include up to ten components, the goodness-of-fit trend does not change significantly after the seven-component model. Consequently, we restrict the number of components shown in Table~\ref{tab:evaluation} from one to seven. As shown in Table~\ref{tab:evaluation}, the five-component model yields the minimum BIC value. Following the guidelines of \citet{Kass&Raftery1995}, the five-component model is decisively favoured over the four-component model ($\Delta\mathrm{BIC} = -27.09$). While the statistical preference for the five-component model over the six-component model is positive rather than decisive ($\Delta\mathrm{BIC} = +5.12$), the estimated physical parameters do not vary significantly between the two. Moreover, the posterior distributions for the five-component model are unimodal, indicating that the fit does not become trapped in a local minimum. Therefore, we adopt the five-component model as our fiducial result. 
The best-fit five-component profiles are displayed in Figure \ref{fig:5component_fit_picture}, with the derived kinematic parameters detailed in Table \ref{tab:5component_fit_parameter}. 
The five-component fit reveals that the strongest \ion{H}{I} absorption is likely a blend of multiple Lyman Limit Systems (LLSs; $10^{17.2} \le \log N_{\mathrm{H\,\textsc{i}}} \ [\mathrm{cm^{-2}}] < 10^{19.0}$ \citep{Tytler+82}) and super-LLSs (SLLSs; $10^{19.0} \le \log N_{\mathrm{H\,\textsc{i}}} \ [\mathrm{cm^{-2}}] < 10^{20.3}$ \citep{Prochaska+04}), with individual column densities ranging from $\log N_{\mathrm{H\,\textsc{i}}} [\mathrm{cm^{-2}}] \sim 18.6$ to $19.9$.
For completeness, the detailed best-fit parameters for the alternative models (1, 2, 3, 4, 6, and 7 components) are summarised in Appendix \ref{app:other_component_results}. 
The full posterior probability distributions and parameter covariances of our MCMC fit using the Python package \texttt{corner} \citep{corner} is presented in Appendix \ref{app:corner_plot}. We caution that the extremely small formal errors on the redshift are artefacts of the restricted $z$ ranges assigned to individual components, and therefore lack physical significance. 

We note that our moderate spectral resolution ($R \sim 1500$) prevents us from definitively resolving the intrinsic line profiles of individual cold clouds. 
While the five-component model provides the most statistically favoured description of the current data, models with a different number of components cannot be entirely ruled out based on these observations alone.

\begin{table}
    \centering
    \caption{The best-fit Voigt profile parameters in the five-component model.}
    \label{tab:5component_fit_parameter}
    \renewcommand{\arraystretch}{1.4}
    \begin{tabular}{cccc} \hline \hline
        Component & $z$ & $\log N_{\text{H\,\textsc{i}}}$ [cm$^{-2}$] & $b$ [km s$^{-1}$] \\ \hline
        1 & $3.000^{+0.0002}_{-0.0002}$ & $18.99^{+0.11}_{-0.17}$ & $22.8^{+11.4}_{-8.9}$ \\
        2 & $3.009$ (fixed)                 & $19.40^{+0.37}_{-0.89}$ & $97.5^{+13.8}_{-22.7}$ \\
        3 & $3.015^{+0.0006}_{-0.0006}$ & $19.92^{+0.11}_{-0.13}$ & $43.8^{+4.9}_{-11.2}$  \\
        4 & $3.025^{+0.0002}_{-0.0003}$ & $18.64^{+0.18}_{-0.34}$ & $25.0^{+11.2}_{-9.3}$ \\
        5 & $3.030^{+0.0002}_{-0.0002}$ & $18.99^{+0.09}_{-0.10}$ & $37.0^{+6.4}_{-11.4}$  \\ \hline
    \end{tabular}
\end{table}

\subsubsection{Metal line identification}
\label{subsubsec:metal_line_identification}
\begin{table}
    \centering
    \caption{Best-fit parameters for the metal absorption lines in Component 2.}
    \label{tab:metal_fit_parameters}
    \renewcommand{\arraystretch}{1.4}
    \begin{threeparttable}
        \begin{tabular}{l c c c} \hline \hline
            Ion & Transition & $\log N \, [\mathrm{cm}^{-2}]$ & $b$ [$\mathrm{km \, s^{-1}}$] \\ \hline
            \ion{O}{I}           & $\lambda$1302              & $17.28^{+0.91}_{-0.78}$  & $68.52^{+15.65}_{-10.64}$ \\
            \ion{Si}{II}         & $\lambda$1304 & $16.06^{+1.73}_{-0.85}$ & $46.11^{+29.58}_{-15.40}$ \\
            \ion{C}{II}          & $\lambda$1334 & $17.23^{+0.52}_{-0.61}$  & $46.62^{+6.16}_{-4.27}$ \tnote{a} \\
            \ion{C}{II*}         & $\lambda$1335 & $14.69^{+0.12}_{-0.11}$  & $46.62^{+6.16}_{-4.27}$ \tnote{a} \\
            \ion{Si}{IV}         & $\lambda\lambda$1393, 1402 & $14.23^{+0.03}_{-0.03}$  & $160.78^{+12.20}_{-11.86}$ \\ \hline
        \end{tabular}
        \begin{tablenotes}
            \footnotesize
            \item[a] The Doppler parameter $b$ of \ion{C}{II*} is tied to that of \ion{C}{II} during the fitting.
        \end{tablenotes}
    \end{threeparttable}
\end{table}

\begin{figure}
    \centering
    \includegraphics[width=\columnwidth]{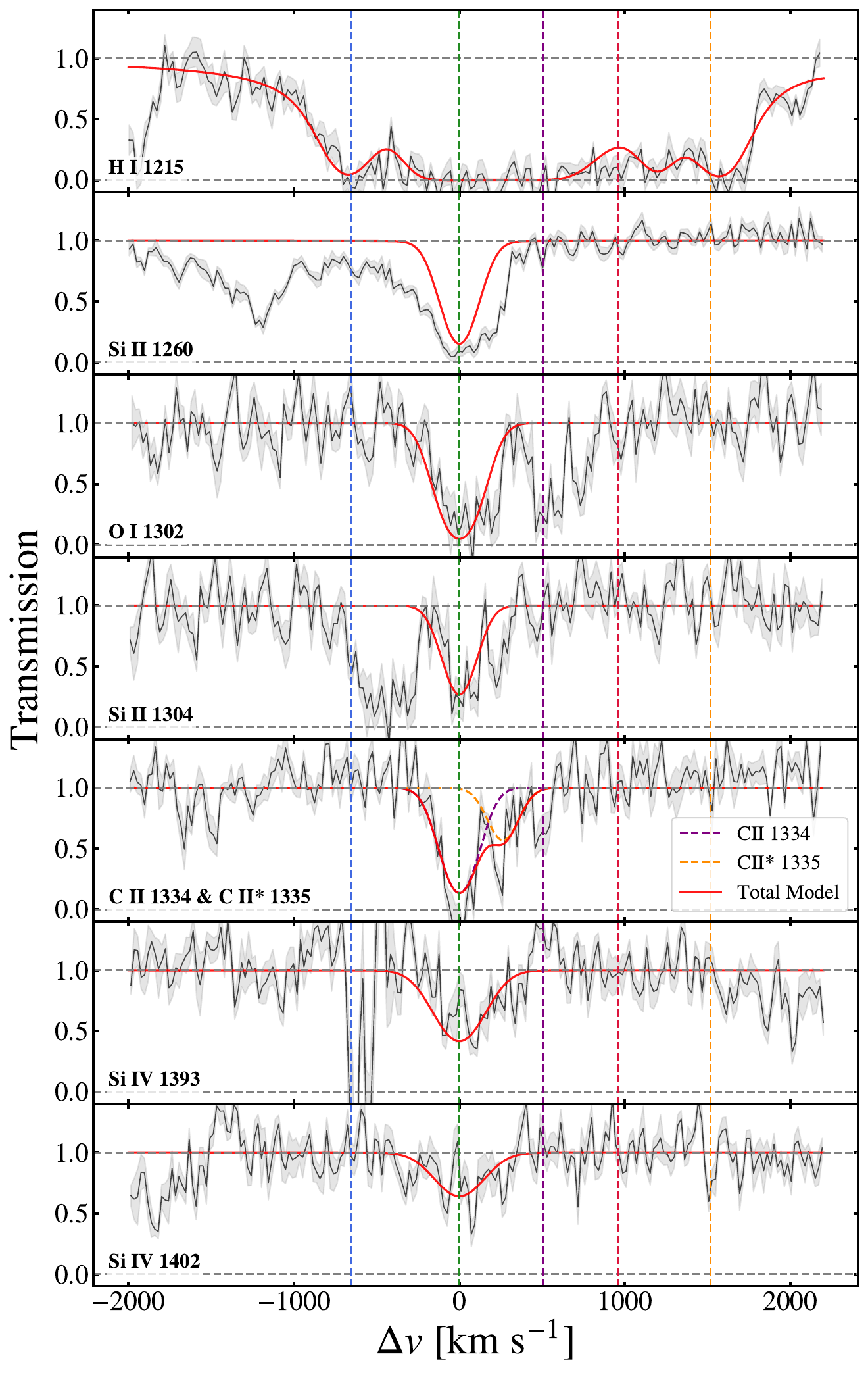}
    \caption{Voigt profile fitting results for the metal absorption lines associated with Component 2 (green vertical line). The transmission spectra are presented in velocity space, centred on the redshift of Component 2. The vertical dashed lines mark the kinematic locations of the five \ion{H}{I} components derived from our fiducial model. Note that the model profile for the \ion{Si}{II} $\lambda 1260$ transition is predicted solely based on the \ion{Si}{II} $\lambda 1304$ fit results.}
    \label{fig:metal_fit_picture}
\end{figure}
Multiple metal transitions associated with Component 2 at $z=3.009$ are found, as shown in Figure \ref{fig:metal_fit_picture}. 
The best-fit Voigt parameters are shown in Table \ref{tab:metal_fit_parameters}.
To obtain these results, we carefully accounted for metal line blending, placing special emphasis on the potential blending affecting the features of Component 2. In the following, we outline the detailed fitting procedure. 

First, the \ion{Si}{ii} $\lambda$1304 line of Component 2 partially overlaps with the \ion{O}{i} $\lambda$1302 transition of Component 3 ($z=3.015$). Nevertheless, the overall metal absorption associated with Component 3 appears to be minimal, showing only a marginal detection of \ion{C}{ii} $\lambda$1334. Furthermore, given that Component 2 exhibits robust multi-transition silicon features (including \ion{Si}{ii} $\lambda$1260, \ion{Si}{iii} $\lambda$1206, and the \ion{Si}{iv} doublet), we assume that the metal contribution from Component 3 to this blended feature is negligible.

Second, we find that the \ion{Si}{ii} $\lambda 1260$ feature of Component 2 is likely blended with the \ion{N}{v} $\lambda\lambda 1238, 1242$ doublet associated with an \ion{H}{I} absorption system at $z \sim 3.06$. Although we find no apparent \ion{C}{IV} absorption in the SDSS spectrum to corroborate this, the high noise level of the data prevents us from definitively ruling out the presence of the \ion{N}{v} doublet. Furthermore, an unphysically broad component appears in this wavelength range, which may arise from poor continuum normalisation caused by masking. Indeed, the uncertainty in the continuum estimation is higher in this region compared to others (Figure~\ref{fig:QFA_fit}). Therefore, we exclude the \ion{Si}{ii} $\lambda1260$ feature from the fit. Moreover, \ion{Si}{ii} $\lambda1526$ falls within the SDSS data that is of relatively poor quality, resulting in a high degree of uncertainty in its identification.
Consequently, we derive the \ion{Si}{ii} column density solely from the \ion{Si}{ii} $\lambda1304$ absorption line.

To ensure a robust analysis and mitigate the degeneracy between the derived column densities and Doppler parameters, we perform simultaneous fits for distinct groups of transitions independently. Specifically, we apply this simultaneous fitting to the \ion{O}{I} $\lambda 1302$ and \ion{Si}{II} $\lambda 1304$ pair, the \ion{C}{II} $\lambda 1334$ and \ion{C}{II*} $\lambda 1335$ pair, and the \ion{Si}{IV} $\lambda\lambda 1393, 1402$ doublet. We ensure that the kinematic parameters ($b$) are tied together for transitions originating from the same ion and slightly adjust the initial parameter ranges to achieve convergence and avoid local minima. For the \ion{O}{I} and \ion{Si}{II} pair, a strictly fixed redshift tied between the two transitions hindered the convergence of the MCMC chains. We therefore relax this condition and allow their redshifts to vary independently within a strictly limited narrow window ($\Delta z \le 0.0005$) to achieve a stable fit.

\subsubsection{Metallicity measurement}
Having robustly measured the column densities, we then perform photoionisation modelling using \texttt{CLOUDY} \citep{C25} to characterise the physical properties of Component 2. 
We constrain the metallicity by incorporating the electron density derived from the \ion{C}{II}/\ion{C}{II*} ratio as an additional constraint alongside $N_{\text{H\,\textsc{i}}}$. In contrast, the metallicities of the other four components remain poorly constrained due to a lack of diagnostics for their ionisation states and the non-detection of metal absorption lines. Nevertheless, we place upper limits on their metallicities based on the 3$\sigma$ detection limits of the corresponding metal transitions.

The electron density, $n_e$, is then calculated from the column density ratio of the \ion{C}{II} ground and excited states (\ion{C}{II} / \ion{C}{II*}) using the following analytical expression:
\begin{equation}
    n_e = n_{\rm cr} \left[ \frac{N_l}{N_u} \left( \frac{g_u}{g_l} \right) e^{-\Delta E/kT} - 1 \right]^{-1},
\end{equation}
where $n_{\rm cr}$ is the critical density, $N_l$ and $N_u$ are the column densities of the ground and excited states, respectively, and $g_l$ and $g_u$ are their corresponding statistical weights ($g_u/g_l = 2$ for the relevant energy levels of \ion{C}{II}). Here, $\Delta E$ represents the energy difference between the ground and excited states, $k$ is the Boltzmann constant, and $T$ is the gas temperature.

For the \ion{C}{II} fine-structure transition, we adopt an energy difference of $\Delta E = 1.26 \times 10^{-14} \, \mathrm{erg}$ and a critical density of $n_{\rm cr} \sim 50 \, \mathrm{cm^{-3}}$, assuming a CGM typical gas temperature of $T=10^4 \, \mathrm{K}$ \citep{Tayal+08, NIST_ASD, Tumlinson+17}. Incorporating these adopted values and our measured column densities, an electron density is derived as $n_e = 0.074^{+0.23}_{-0.05} \, \mathrm{cm^{-3}}$ for Component 2, where the uncertainties denote the $1\sigma$ confidence intervals estimated from the MCMC posterior distributions. Although we vary the temperature in the calculation, the derived electron density changes only slightly because the temperature dependencies of the critical density and the other factor cancel each other out. 
For our photoionisation modelling with \texttt{CLOUDY}, we adopt a constant total hydrogen density of $n_{\mathrm{H}} = 10^{-1.2} \, \mathrm{cm^{-3}}$. It is derived from the measured electron density using the relation $n_e \approx 1.2 \, n_{\mathrm{H}}$, which is a standard approximation for highly ionised gas with primordial or solar helium abundances. Our \texttt{CLOUDY} model \textit{a posteriori} validates this assumption; the resulting ionisation structure demonstrates that the cloud is highly ionised with an \ion{H}{II} fraction of $\sim 90\%$, and the predicted electron density in the highly ionised zones ($n_e \approx 0.07 \, \mathrm{cm^{-3}}$) is consistent with our measurement. Assuming the cosmic ultraviolet background at $z=3.009$ \citep{FG20}, our \texttt{CLOUDY} model yields an ionised hydrogen column density of $N_{\mathrm{H\,\textsc{ii}}} = 10^{20.51} \, \mathrm{cm^{-2}}$. By comparing column densities derived from this model to our observed column densities, we calculate the intrinsic metallicity of the absorption system. The derived metallicities are summarised in Table \ref{tab:metallicity_results}. Notably, the carbon, oxygen and silicon abundances exhibit super-solar values. Overall, the derived column densities and Doppler parameters represent the most robust constraints achievable with the current observational data. 

To constrain the metallicities of the remaining \ion{H}{I} components with no detections of metal absorptions, we estimate the $3\sigma$ upper limits on their rest-frame equivalent widths and convert them into column density upper limits. First, the total $1\sigma$ uncertainty of the equivalent width, $\sigma_{\rm EW}$, is calculated by adding the statistical flux uncertainty ($\sigma_{\mathrm{flux}}$) and the continuum placement uncertainty ($\sigma_{\mathrm{cont}}$) in quadrature: $\sigma_{\mathrm{EW}} = \sqrt{\sigma_{\mathrm{flux}}^2 + \sigma_{\mathrm{cont}}^2}$. Following \citet{Chen+15}, these two terms are evaluated as:
\begin{align}
    \sigma_{\mathrm{flux}} &= \frac{\sqrt{\sum_i P^2 (\lambda_i-\lambda_0) \sigma_{\mathrm{f}_i}^2}}{\sum_i P^2 (\lambda_i-\lambda_0)} \Delta \lambda_{\mathrm{pix}}, \\
    \sigma_{\mathrm{cont}} &= \frac{A(\lambda_{\mathrm{max}}-\lambda_{\mathrm{min}})}{\mathrm{SNR}},
\end{align}
where $P(\lambda_i - \lambda_0)$ represents the assumed line profile centred at the rest wavelength $\lambda_0$, $\lambda_i$ is the pixel wavelength, $\Delta \lambda_{\mathrm{pix}}$ is the wavelength step per pixel, and $\sigma_{\mathrm{f}_i}$ is the normalised $1\sigma$ flux uncertainty at each pixel, incorporating both the statistical data error and the estimated intrinsic spectrum error. Although QFA provides an estimate for the intrinsic spectrum error, \textsc{SpenderQ} does not. Consequently, following \citet{SpenderQ}, which reports a median absolute fractional flux error of 0.04, we adopt a constant fractional error of 0.04 for the \textsc{SpenderQ} predictions. 
Here, SNR represents the local S/N, calculated as the ratio of the mean continuum flux to the median flux error within the adjacent unabsorbed regions. We adopt an amplitude factor of $A=0.5$ \citep{Misawa+14} and set the integration window to $\lambda_{\mathrm{max}} - \lambda_{\mathrm{min}} = 6\sigma_{\mathrm{line}}$, where $\sigma_{\mathrm{line}}$ is the Gaussian line width derived from the fits to the detected metal lines.
For these velocity components lacking statistically significant metal absorption features, the $3\sigma$ rest-frame equivalent width upper limit, $EW_\mathrm{lim}$, can be formulated as $EW_\mathrm{lim} = 3\sigma_{\mathrm{EW}} /(1+z)$.
Assuming these non-detected transitions are optically thin and thus reside on the linear part of the curve of growth, we convert the $EW_\mathrm{lim}$ into column density upper limits, $N_{\mathrm{lim}}$, using the following standard relation \citep{Draine+11}:
\begin{equation}
    N_{\mathrm{lim}} = \frac{m_e c^2}{\pi e^2} \frac{EW_{\mathrm{lim}}}{f\lambda_0^2},
\end{equation}
where $f$ is the oscillator strength of the transition, $\lambda_0$ is its rest-frame wavelength, and $m_e$, $c$, and $e$ represent the electron mass, the speed of light, and the elementary charge, respectively.  
Table \ref{tab:combined_column_densities} shows the upper limits of column densities and the corresponding upper limits of metallicities for the four components for which no metal lines are detected.
It can be seen that all four components have significantly lower metallicities compared to Component 2.
In summary, this analysis demonstrates a chemical contrast: these four surrounding components exhibit definitively low metallicities, in sharp contrast to Component 2, which is the primary metal-rich absorber at $z=3.009$.

\begin{table}
    \centering
    \setlength{\tabcolsep}{4pt}
    \caption{Estimated metallicities and ionization corrections for Component 2}
    \begin{threeparttable}
        \renewcommand{\arraystretch}{1.4}
        \begin{tabular}{l c c c c}
            \hline \hline
            Element & Ion & $\log N_{\mathrm{obs}}$ & $\log N_{\mathrm{model}}$ & $[\mathrm{X/H}]$ \\
            & & $[\mathrm{cm}^{-2}]$ & $[\mathrm{cm}^{-2}]$ & [dex] \\ \hline
            Oxygen  & \ion{O}{I}   & $17.28^{+0.91}_{-0.78}$ &$16.088$ & $+1.19^{+0.91}_{-0.78}$ \\
            Carbon  & \ion{C}{II}  & $17.23^{+0.52}_{-0.61}$ & $16.097$ & $+1.13^{+0.52}_{-0.61}$ \\
            Silicon & \ion{Si}{II} & $16.06^{+1.73}_{-0.85}$ &  $15.463$ & $+0.60^{+1.73}_{-0.85}$ \\
            \hline
        \end{tabular}
    \end{threeparttable}
    \label{tab:metallicity_results}
\end{table}

\begin{table*}
    \centering
    \caption{Summary of column densities ($\log N \, [\mathrm{cm}^{-2}$]) and estimated metallicities for the five-component model. The values indicated by the inequality sign are the $1\sigma$ upper limit.}
    \label{tab:combined_column_densities}
    \renewcommand{\arraystretch}{1.4}
    \begin{tabular}{lccccc}
        \hline \hline
        Component & 1 & 2 & 3 & 4 & 5  \\ \hline
        $z$ & $3.000$ & $3.009$ & $3.015$ & $3.025$ & $3.030$ \\
        \hline
        $\log N_{\text{H\,\textsc{i}}}$   & $18.99^{+0.11}_{-0.17}$ & $19.40^{+0.37}_{-0.89}$ & $19.92^{+0.11}_{-0.13}$ & $18.64^{+0.18}_{-0.34}$ & $18.99^{+0.09}_{-0.10}$ \\
        $\log N_{\text{O\,\textsc{i}}}$   & $< 14.85$ & $17.28^{+0.91}_{-0.78}$ & $< 14.85$ & $< 14.85$ & $< 14.87$ \\
        $\log N_{\text{C\,\textsc{ii}}}$  & $< 14.28$ & $17.23^{+0.52}_{-0.61}$ & $< 14.28$ & $< 14.28$ & $< 14.30$ \\
        $\log N_{\text{Si\,\textsc{ii}}}$ & $< 14.59$ & $16.06^{+1.73}_{-0.85}$ & $< 14.59$ & $< 14.62$ & $< 14.60$ \\
        \hline
        $[\mathrm{O/H}]^a$        & $< -0.66$ & $+1.19^{+0.91}_{-0.78}$ & $< -1.63$ & $< -0.14$ & $< -0.71$ \\
        \hline
    \end{tabular}
    \vspace{1ex}
    \begin{minipage}{\textwidth}
        \footnotesize
        $^a$ Estimated apparent metallicity assuming $[\mathrm{O/H}] \approx [\text{O\,\textsc{i} / H\,\textsc{i}}]$ with a solar abundance of $\log(\mathrm{O/H})_\odot = -3.31$ \citep{Asplund+21}.
    \end{minipage}
\end{table*}

While the five-component model is optimal for our current dataset, we acknowledge that the limited spectral resolution ($R\sim1500$) introduces uncertainty regarding the exact number of kinematic components. Furthermore, in the simultaneous fitting of the Ly$\alpha$ and Ly$\beta$ transitions, the result heavily depends on the Ly$\alpha$ profile due to the relatively poor S/N in the Ly$\beta$ region. 
To account for these uncertainties and ensure the robustness of our chemical abundance estimates, we also investigate the derived metallicities under alternative component numbers.  
Specifically, we explore a two-component scenario. 
Reducing the number of kinematic components inherently assigns a larger fraction of the total \ion{H}{I} column density to the primary absorption system. This increased $N_{\text{H\,\textsc{i}}}$ denominator would, in turn, lower the inferred metallicity to a more typical and realistic value for an absorber at this redshift. 
To test this, we evaluate a two-component kinematic model, where the \ion{H}{I} column density assigned to the $z=3.009$ system increases to $N_{\text{H\,\textsc{i}}} = 10^{20.37} \, \mathrm{cm^{-2}}$. The corresponding \texttt{CLOUDY} model yields 
the metallicities $[\mathrm{O/H}] = +0.20$, $[\mathrm{C/H}] = +0.45$, and $[\mathrm{Si/H}] = -0.02$. 
It can be seen that even when the metallicity is intentionally minimised on this highly conservative kinematic assumption, the system maintains super-solar metallicities at least for oxygen and carbon. 
However, to draw a definitive conclusion regarding the super-solar nature of this system, higher-resolution spectroscopic data and multi-wavelength observations will likely be required in the future.

\subsubsection{Stacking analysis of the background galaxies}
\begin{figure}
    \centering
    \includegraphics[width=\columnwidth]{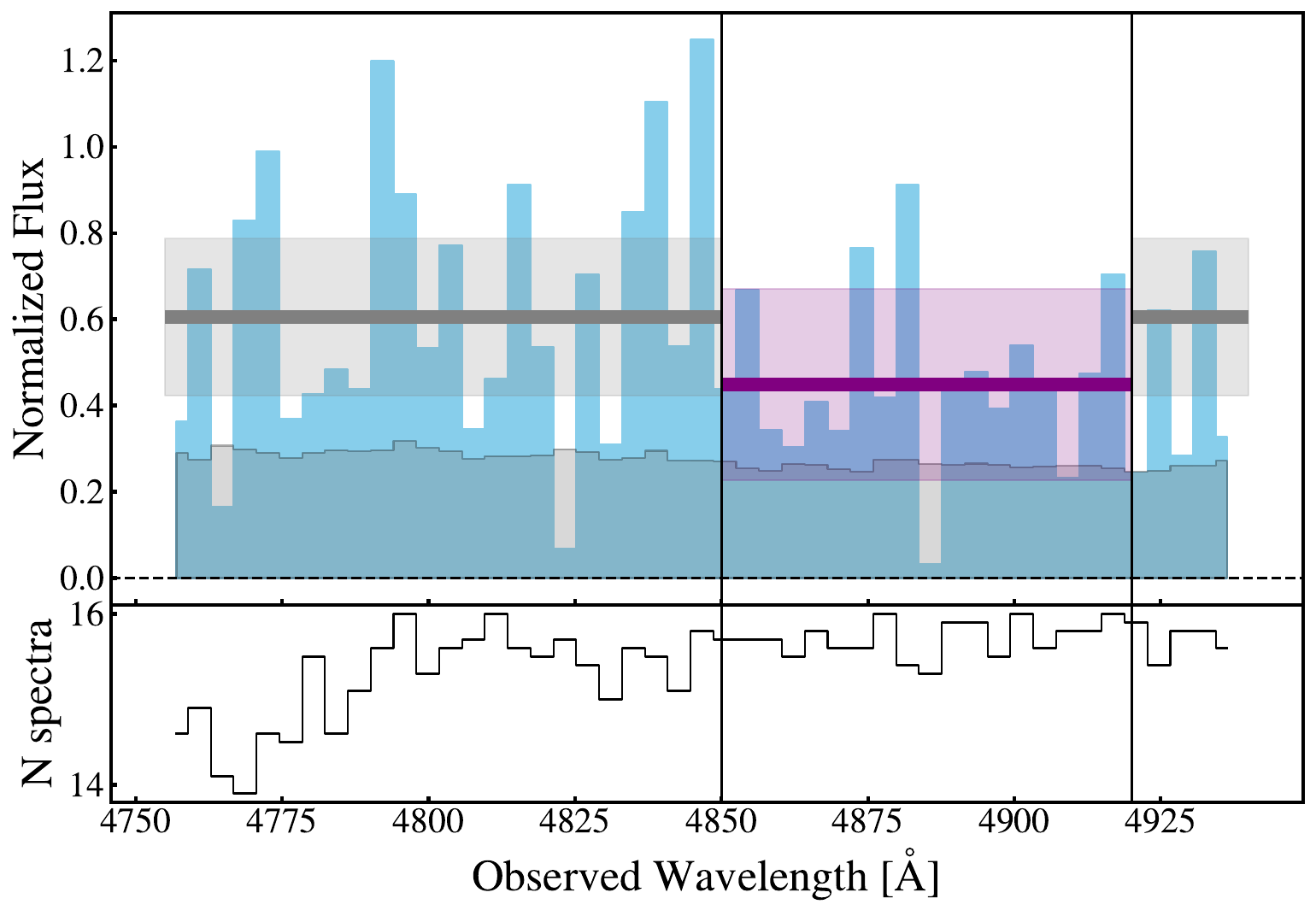}
    \caption{Median stacked spectrum of the background galaxies ($N=19$). The vertical black lines indicate the redshift range corresponding to the strongest \ion{H}{I} absorption. The shaded grey region represents the $1\sigma$ uncertainty estimated from bootstrap resampling. The broad horizontal lines denote the mean flux levels in the corresponding wavelength regions; the strong absorption region is shown in purple, and outside regions are shown in grey. For visualization purposes, the spectrum has been binned by 10 pixels. The lower panel shows the histogram of the number of spectra used for stacking at each wavelength, which has also been rebinned into the same 10-pixel bins.}
    \label{fig:DLA_stack}
\end{figure}

We stack the spectra of the background sources located behind the strongest \ion{H}{I} absorption system, just in case we might be able to detect a coherent absorption signature.
Following a procedure similar to the protocluster stacking analysis described in Section \ref{subsec:HIgas_PC}, we initially select a sample of 23 background galaxies, represented by the triangles and squares in Figure \ref{fig:sky_dist_relative}, against the strongest \ion{H}{I} absorption. As mentioned previously, four sources are excluded due to insufficient spectral coverage or negative normalisation factors at rest-frame $1280$\,\AA, leaving a final sample of 19 galaxies. The global stack encompassing all sightlines yields a flux difference between the strongest \ion{H}{I} absorption region and the non-absorption region of $0.14 \pm 0.10$. The significance of this difference is therefore $1.4\sigma$. Although the flux difference has the positive sign, suggesting a possible enhancement of \ion{H}{I} absorption in this region, the current sample size is insufficient to establish this trend with statistical significance. Figure \ref{fig:DLA_stack} shows the result of the stacking analysis, where the symbols are used in the same way as in Figure \ref{fig:stacked_PC}.
We further stack the spectra of the background galaxies with small impact parameters relative to the quasar; however, no significant absorption signal is detected.
Deeper observations or a larger sample of background galaxies will be required to draw a definitive conclusion. 
Several factors may account for this null detection, including the limited S/N of the background galaxy spectra and the possibility that the absorber is intrinsically compact in its transverse physical extent.

\subsubsection{The possibility of a BAL QSO}
Finally, we consider the possibility that this massive absorption feature is intrinsic to the background quasar itself. 
Although the extreme velocity width of our primary \ion{H}{I} absorption trough ($> 2,000 \, \mathrm{km \, s^{-1}}$) is characteristic of a typical Broad Absorption Line (BAL) feature of a quasar \citep[e.g.][]{Murray+95, Elvis+00}, the system lacks the expected associated high-ionisation metal lines. While the \ion{C}{IV} $\lambda\lambda$1548, 1550 doublet falls outside our Subaru/FOCAS spectral coverage, inspection of the archival SDSS spectrum reveals an absence of broad \ion{C}{IV} absorption---the defining signature of BAL quasars. Furthermore, our measurements of the Balnicity index \citep[BI;][]{Weymann+91} and the absorption index \citep[AI;][]{Hall+02} both yield values of $0 \, \mathrm{km \, s^{-1}}$, even when measured on a spectrum smoothed with a five-pixel moving average. Finally, the official SDSS data release catalogue \citep{SDSS_DR17} does not flag this object as a BAL quasar. We therefore rule out an intrinsic BAL origin, concluding that this \ion{H}{I} complex is an intervening foreground structure. 

\section{Discussion}
\label{sec:discussion}
\subsection{Absence of strong \texorpdfstring{\ion{H}{I}}{HI} absorption in the protocluster}
\label{subsec:absence_HI}
We find no strong \ion{H}{I} absorption in the quasar spectrum within the redshift range of the protocluster (Figure \ref{fig:transmission_galaxy}). Furthermore, the stacked spectrum of the background sources similarly yields a null result (Figure \ref{fig:stacked_PC}). 
To interpret this absence of \ion{H}{I} gas, we evaluate how much absorption should be expected if the IGM simply traces the underlying mass distribution. 
We calculate the mass overdensity, $\delta_m$, of the protocluster. First, to facilitate a direct comparison with the cosmological simulations presented in \citet{Miller+19}, we scale our observed galaxy overdensity, $\delta_{\mathrm{g}} = 2.49$, to their adopted volume of $(15 \, h^{-1} \mathrm{cMpc})^3$. The member galaxies in our protocluster extend over approximately $5 \, h^{-1} \mathrm{cMpc}$, $10 \, h^{-1} \mathrm{cMpc}$, and $32.7 \, h^{-1} \mathrm{cMpc}$ along the RA, Dec, and LOS directions ($\Delta z=0.05$), respectively, yielding a volume of $1635 \, (h^{-1} \mathrm{cMpc})^3$. Assuming the surrounding field has an average density, the scaled galaxy overdensity within the $(15 \, h^{-1} \mathrm{cMpc})^3$ volume becomes $\delta_{\mathrm{g}} = 1.21$. 
Next, we convert $\delta_\mathrm{g}$ to the underlying mass overdensity, $\delta_\mathrm{m}$, using the effective linear galaxy bias, $b$. 
The spectroscopically confirmed members of our protocluster are predominantly LAEs, and according to previous studies, a typical bias at $z \sim 3$ is $b \approx 2$ for LAEs \citep[e.g.][]{Gawiser+07, Ouchi+10, Ouchi+18}. 
Consequently, we obtain $\delta_\mathrm{m} = \delta_\mathrm{g} / b = 0.61$. 
To verify the robustness of this mass estimation against uncertainties in the assumed galaxy bias, we perform an independent consistency check using the empirical $\delta_\mathrm{g}$--$\delta_\mathrm{m}$ relation derived by \citet{Chiang+13} based on the Millennium Simulation \citep{Springel+05} and a semi-analytic galaxy formation model \citep{Guo+11}. This approach allows us to map our observed galaxy overdensity directly to the underlying mass overdensity, based on their simulated relation, without assuming a linear bias factor. 
Scaling our measured galaxy overdensity to their volume $(15\ \mathrm{cMpc})^3$ yields $\delta_\mathrm{g} = 3.52$; when evaluated using the $z=3$ relation, the mass overdensity is estimated to be $\delta_\mathrm{m} \approx 1.7$.
Converting this value back to our fiducial $(15 \, h^{-1} \mathrm{cMpc})^3$ scale results in $\delta_\mathrm{m} = 0.58$, which shows an agreement with that assuming the bias parameter.

Finally, based on the $\delta_\mathrm{m}$--$\delta_{\tau_{\mathrm{eff}}}$ relation derived from the simulations in \citet{Miller+19}, a mass overdensity of $\delta_\mathrm{m} = 0.61$ corresponds to an expected effective optical depth overdensity of $\delta_{\tau_{\mathrm{eff}}} \approx 1$. According to \citet{Becker+13}, the mean effective optical depth of the IGM evolves with redshift as follows:
\begin{equation}
    \langle \tau_{\mathrm{eff}} (z) \rangle = 0.00958 \times (1+z)^{2.90} - 0.132.
\end{equation}
At $z = 3.079$, the mean effective optical depth is $\langle \tau_{\mathrm{eff}} \rangle = 0.43$. 
Therefore, when $\delta_{\tau_{\mathrm{eff}}} \approx 1$, the expected optical depth of the protocluster is $\tau_{\mathrm{eff}} \approx 0.86$. 
On the other hand, the observed mean transmission within the protocluster redshift range is $0.808\pm0.004$, which corresponds to an effective optical depth of $\tau_{\mathrm{eff}}=0.214 \pm 0.005$, a value smaller than expected. However, given the substantial scatter in the $\delta_\mathrm{m}$--$\delta_{\tau_{\mathrm{eff}}}$ relation reported by \citet{Miller+19}, it is not unexpected that this protocluster lacks strong \ion{H}{I} absorption. 
To determine whether the protocluster is associated with \ion{H}{I} absorption, a larger sample of background sources observed via narrow-band imaging or spectroscopy is required \citep{Mawatari+17, Hayashino+19}. 

\subsection{Comparison of the strongest \texorpdfstring{\ion{H}{I}}{HI} absorption with previous studies}
\begin{figure}
    \centering
    \includegraphics[width=\columnwidth]{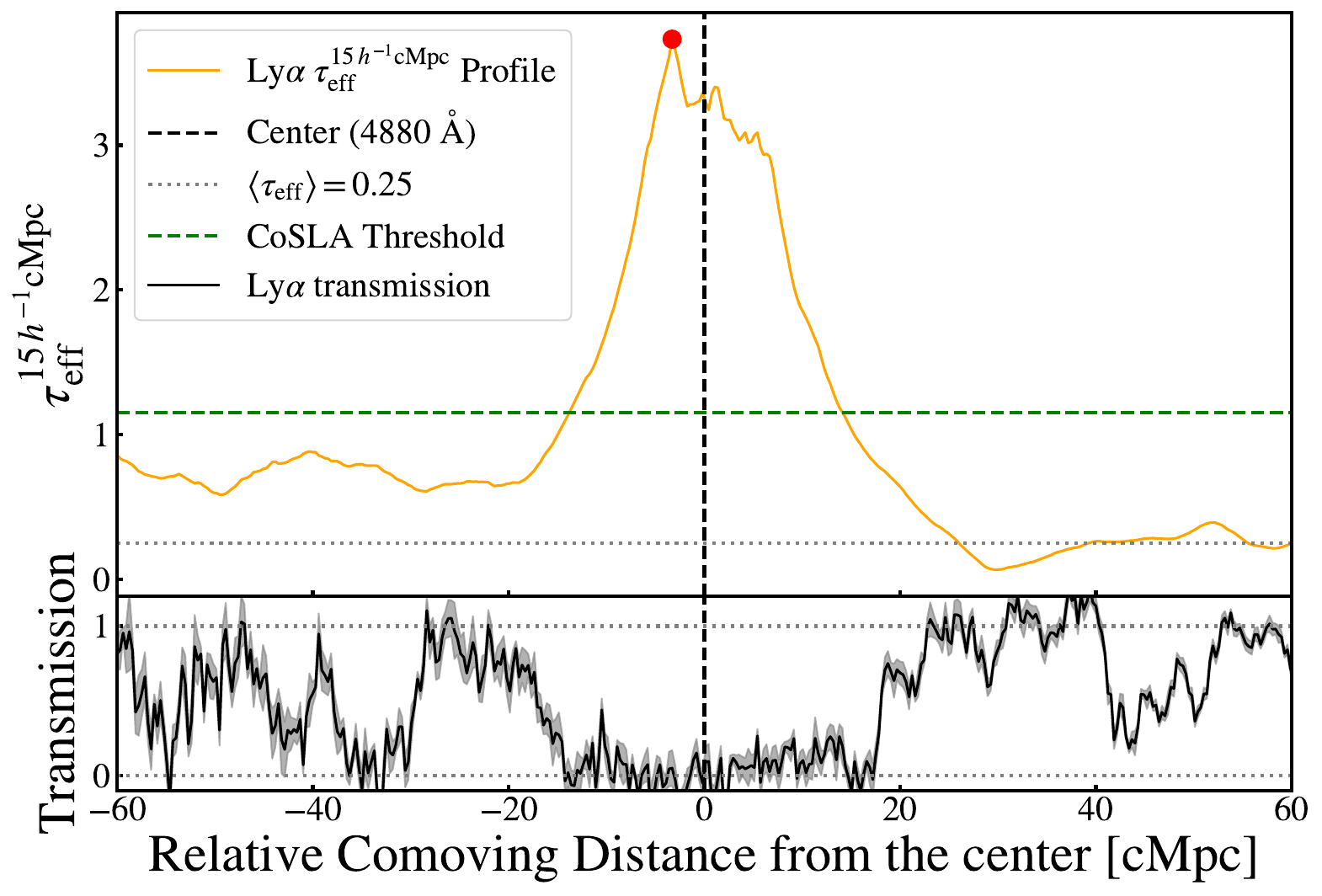}
    \caption{
    Upper panel: Effective optical depth on a $15 \, h^{-1} \, \mathrm{cMpc}$ scale (orange line) for the strongest \ion{H}{I} absorption, with the peak value marked by a red point. The green dashed line represents the CoSLA criterion \citep{Cai+16}, whereas the grey dotted line represents the mean optical depth $\langle \tau_{\mathrm{eff}} \rangle = 0.25$.
    Lower panel: Normalized transmission spectrum of the Ly$\alpha$ transition (black line) and the corresponding uncertainty (grey shaded region). Both panels share a common horizontal axis centred on the \ion{H}{I} absorption trough at $4880~\text{\AA}$.}
    \label{fig:tau_eff}
\end{figure}

This study demonstrates that the observed strongest absorption line, which extends over 40 cMpc, is best explained by a superposition of five gas clouds.
Such a strong absorption feature is characteristic of the Coherently Strong Ly$\alpha$ Absorption (CoSLA) systems, which are identified by \citet{Cai+16} through a systematic search of $\sim 6000$ sightlines from the SDSS quasar survey at $z = 2.6\text{--}3.0$ over a volume exceeding $1 \, (h^{-1} \, \mathrm{Gpc})^3$. 
They assume that a CoSLA exhibits an effective optical depth of $\tau_\mathrm{eff}^{15 h^{-1} \, \mathrm{cMpc}} > 1.15$ on a $15 \, h^{-1} \, \mathrm{cMpc}$ scale, which is substantially higher than the mean intergalactic optical depth ($\langle\tau_{\mathrm{eff}}\rangle \approx 0.25$) at the same redshift. 
Figure \ref{fig:tau_eff} presents the effective optical depth at each LOS position when the strongest absorption line identified in this study are smoothed over $15 \, h^{-1} \, \mathrm{cMpc}$ scale they applied.
The maximum effective optical depth is measured to be $\tau_\mathrm{eff}^{15 h^{-1} \, \mathrm{cMpc}} = 3.73$, which
is more than three times the CoSLA threshold.
This implies that \ion{H}{I} system found in this study is exceptionally massive and extreme, even when compared to the robust absorption complexes typically associated with known large-scale overdensities. 
\citet{Prochaska+06} reported two LLSs at $z \sim 2$ that exhibit super-solar gas-phase metallicities ($\mathrm{[M/H]} = +0.7 \pm 0.2, +0.05 \pm 0.1$ dex), and one of these systems lacks damping wings, closely resembling our strongest \ion{H}{i} absorption feature. 
Combined with our results at $z \sim 3$, these findings suggest that such metal-rich SLLSs may serve as major reservoirs of heavy elements in the early Universe.

\subsection{The origin of the strongest \texorpdfstring{\ion{H}{I}}{HI} absorption}
\begin{figure}
    \centering
    \includegraphics[width=\columnwidth]{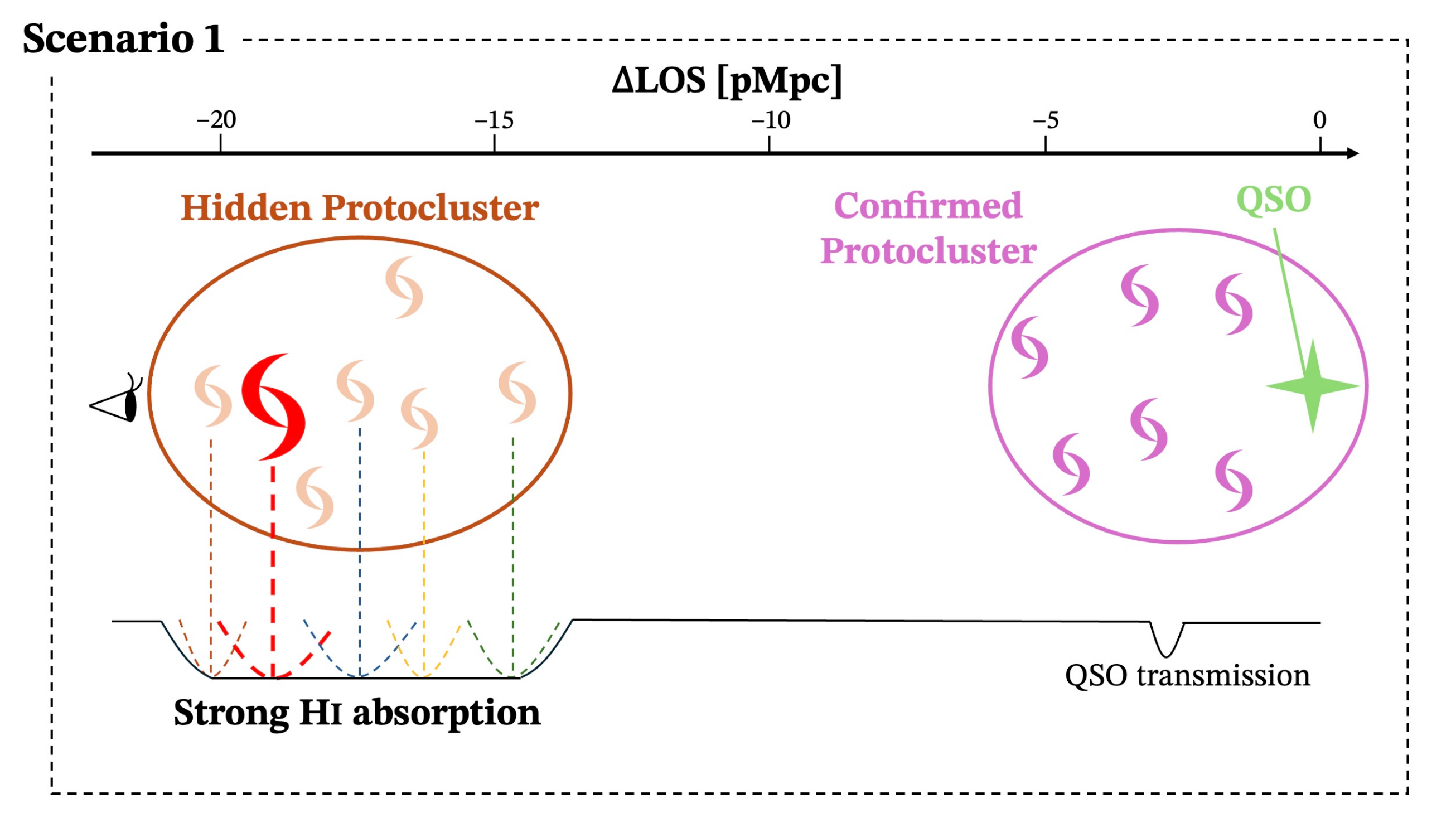}
    \includegraphics[width=\columnwidth]{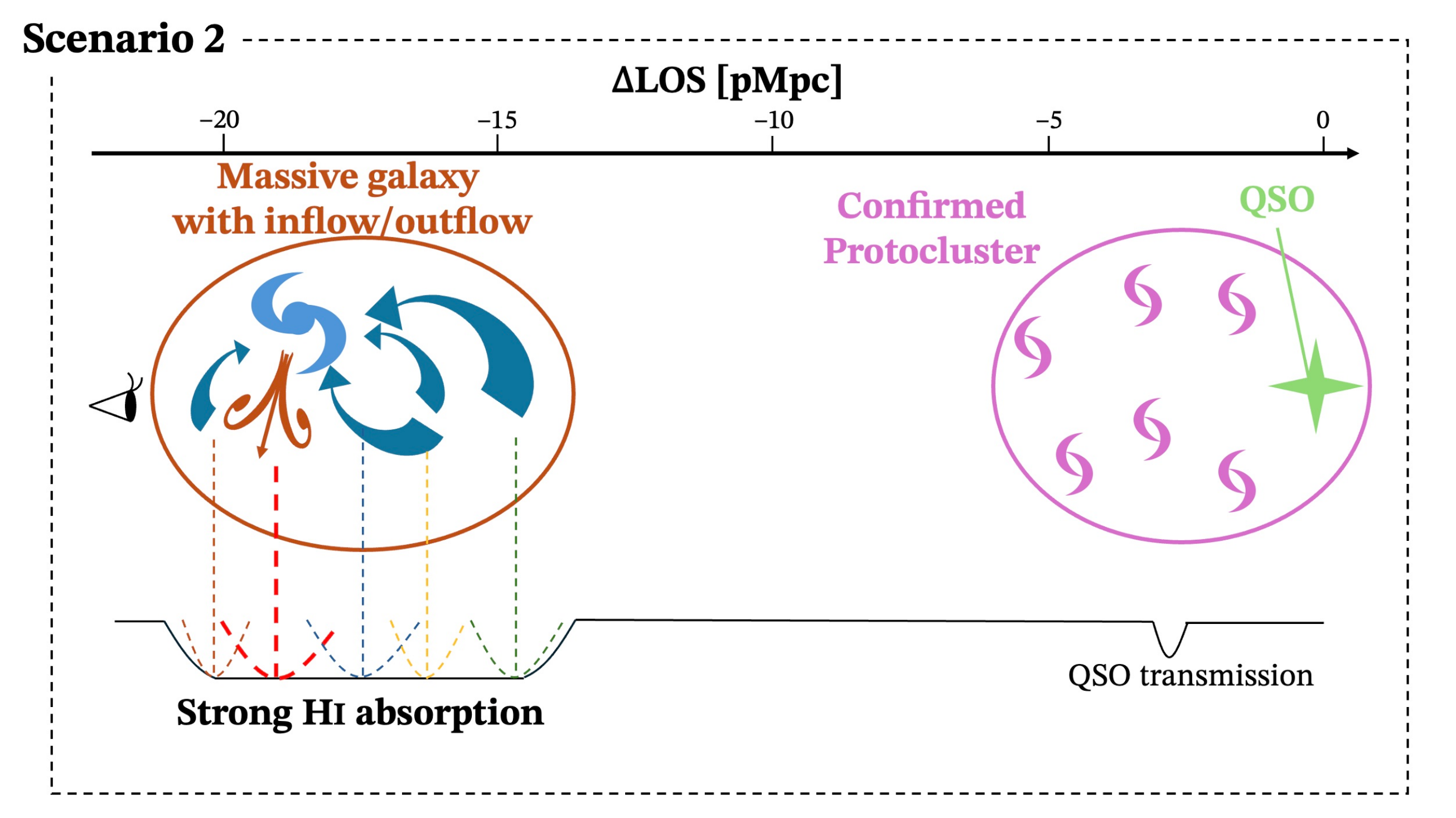}
    \caption{Schematic illustrating the two proposed physical scenarios for the strongest \ion{H}{I} absorption system. Each panel features a $\Delta\mathrm{LOS}$ axis at the top, with the origin set at the quasar position, and illustrates the corresponding quasar transmission profile at the bottom. 
    Upper panel: The hidden protocluster scenario, in which the strong absorption profile results from the kinematic superposition of the circumgalactic media (CGM) surrounding multiple, unobserved member galaxies along the LOS. 
    Lower panel: The single-halo scenario, depicting metal-rich galactic outflows and pristine or metal-poor inflows driven by a single massive, metal-rich host galaxy.}
    \label{fig:two_scenario}
\end{figure}

Our analysis reveal a unique set of properties for the strongest \ion{H}{I} absorption: it exhibits a complex five-component kinematic structure, one of which exhibits an super-solar metallicity, yet 
no corresponding galaxies have been identified. 
To naturally account for these seemingly contrasting properties-a highly metal-enriched one gas clump within multiple gas clumps that lacks a UV-bright host counterpart-we propose two possible physical scenarios. Figure \ref{fig:two_scenario} shows the two scenarios: a hidden protocluster and a massive galaxy associated with inflow/outflow.

\subsubsection{Hidden protocluster}
In the first scenario, we consider the presence of an additional, hidden massive protocluster or large-scale structure at $z \sim 3.01$. As demonstrated by our metal absorption analysis, this system contains a heavily metal-enriched component (Component 2). 
Assuming this originated in one or more chemically evolved galaxies, this massive \ion{H}{I} absorption may arise from the kinematic superposition of the CGMs of multiple member galaxies intersecting our LOS. 
The apparent absence of these member galaxies in our current data can be explained by a combination of observational biases and dust obscuration. Primarily, our spectroscopy strategy is optimised for identifying the protocluster members and background galaxies behind the quasar ($z =3.09$). By applying a $z_{\mathrm{phot}}$ cut to select sources behind the quasar, faint LBGs associated with this foreground $z \sim 3.01$ system might be systematically excluded from our spectroscopic sample, despite the $z_{\mathrm{phot}}$ uncertainty. In fact, within an impact parameters of $< 2\,\mathrm{cMpc}$ from the quasar LOS, there are seven unobserved $U$-dropout galaxies whose redshifts are unknown. 

Alternatively, the host galaxies might be dusty star-forming galaxies (DSFGs), whose severe dust obscuration renders them invisible in the rest-frame UV. 
To evaluate the plausibility of the dusty environment, we examine the dust extinction in the intrinsic rest-frame UV continuum of the background quasar. We conduct a power-law fit to the intrinsic quasar continuum, which is estimated using QFA at wavelengths redward of the Ly$\alpha$ emission. The derived UV continuum slope is $\beta = -1.515$, which is consistent with the canonical quasar value \citep[$\beta = -1.56$;][]{VandenBerk+01}. This apparent lack of continuum reddening does not support the presence of heavy dust extinction in front of the quasar. However, if the dense dust obscuring DSFGs is localized to the galaxies themselves, the quasar sightline could pass through their extended, metal-enriched CGM, effectively avoiding the dust-shrouded central regions.

A remaining challenge to this protocluster scenario, however, is the total velocity width of the \ion{H}{I} absorption feature, which extends to approximately $\pm 2000 \, \mathrm{km \, s^{-1}}$. This kinematic spread significantly exceeds the typical velocity widths of virialized clusters or standard CoSLA systems.
Rather than a compact, spherically virialized cluster, we might be observing structures,
such as a cosmic filament or a violently merging galaxy group, aligned directly along the LOS. 
Assuming such an elongated configuration naturally explains the extreme apparent velocity width along the sightline; furthermore, since we can expect the absorbing gas to be confined to the vicinity of the sightline, this is consistent with the null results of our stacking analysis of background galaxies (Section \ref{subsec:HIgas_PC}). 

Indeed, this interpretation aligns with recent discoveries by LATIS of enigmatic structures characterised by exceptionally strong \ion{H}{i} absorption yet lacking any detectable luminous galactic counterparts at $z \sim 2.5$ \citep{Newman+25}. 
They discover a UV-dim protocluster candidate, devoid of significant LAE/LBG overdensities or submillimeter-selected sources. However, unlike this structure, our strongest \ion{H}{i} absorber exhibits associated metal absorption, which could provide a clue to unravelling the nature of these UV-dim overdensities. In contrast, \citet{Dong+23} reported a $z=2.30$ protocluster that lacks significant \ion{H}{i} absorption, interpreting this deficit as the result of large-scale AGN jet feedback or early gravitational shock heating. Therefore, our metal-rich, strong \ion{H}{i} absorber presents an intriguing paradox. 
While the super-solar metallicity suggests a highly evolved system that has undergone significant chemical enrichment, the massive reservoir of cold \ion{H}{i} gas implies that the structure is still in an early, unheated stage of its physical evolution. 

\subsubsection{Outflow and inflow associated with a massive galaxy}
Alternatively, we consider a second scenario wherein the absorption complex traces multi-phase gas flows associated with a single massive host galaxy. Within this framework, the one metal-bearing component is interpreted as a highly enriched outflow or a CGM of a single massive galaxy, likely driven by intense starburst or AGN feedback. Conversely, the remaining four metal-poor components would represent pristine or metal-poor gas inflows accreting from the surrounding IGM. 

However, a severe physical challenge undermines this single-halo model. The total velocity spread across the five components extends over approximately $2000 \, \mathrm{km \, s^{-1}}$. It is highly implausible that gas exhibiting such an extreme velocity dispersion could remain gravitationally bound to a single galactic halo. Furthermore, if the metal-poor components indeed represent infalling gas, their inward peculiar velocities would oppose the cosmological Hubble expansion along the LOS. Consequently, the observed velocity spread actually underestimates the true physical separation of the gas clouds. Accounting for this kinematic compression, the intrinsic spatial scale of the structure must be larger than what the naive Hubble flow conversion suggests, placing it firmly in the protocluster-scale regime ($\gtrsim 10 \, \mathrm{pMpc}$) rather than a single circumgalactic environment. 

Therefore, the reality likely lies in a hybrid scenario, combining elements of both models. The system could represent violent, metal-rich outflows interacting with multiple merging galaxies within a dense protocluster environment.

\section{Conclusions}
\label{sec:conclusions}
In this study, we investigate a unique system - an extremely rare configuration in which a protocluster candidate lies in front of a background quasar at $z=3.09$.
Our main results are summarized as follows:

\begin{enumerate}
    \item A protocluster identification: Our spectroscopic observation detects 42 LAEs, one LBG and brings high-S/N spectrum of the quasar. Among these, 12 members constitute a protocluster at $z=3.079$ with a significant overdensity $\delta = 2.49$ that spatially encompasses the quasar sightline.

    \item Absence of \ion{H}{I} absorption in the protocluster: We observe a lack of strong, coherent \ion{H}{I} absorption in the quasar spectrum within the protocluster redshift range. While the expected \ion{H}{I} effective optical depth is $\tau_\mathrm{eff} \approx 0.86$, the observed mean optical depth is lower ($\tau_\mathrm{eff} = 0.214 \pm 0.005$). However, the expected $\tau_\mathrm{eff}$ exhibits a large scatter, making it difficult to conclude whether this protocluster contains a sufficient amount of \ion{H}{I}.
    
    \item The strongest \ion{H}{I} absorber: The most prominent \ion{H}{I} absorption feature with velocity width of $\sim 2000$ km s$^{-1}$ ($\sim40$ cMpc) in the background quasar spectrum is located at $z \sim 3.01$ and does not correspond to the main $z = 3.079$ protocluster. This system could be resolved into a complex five-component structure with column densities ranging from $\log N(\text{H\,\textsc{i}}) = 18.64$ to $19.92 \, \mathrm{cm^{-2}}$.
    
    \item Extreme chemistry and localized nature: The \texttt{CLOUDY} modelling reveals that one of the five components possesses a super-solar metallicity of $[\mathrm{O/H}] \approx +1.2$. However, the other components have much lower metallicities.
    
    \item Physical origins: We propose two physical scenarios for this unique absorber: (1) another massive, hidden protocluster intersecting the LOS, or (2) a circumgalactic structure associated with a single massive galaxy with metal-rich outflows and metal-poor inflows from the IGM.
\end{enumerate}

These results are achieved by examining both the ``shadow'' of absorption lines appearing in the quasar spectrum and the ``light'' of the corresponding galaxies detected by spectroscopic observation. 
While \ion{H}{I} absorption lines without galaxy counterparts have been observed in the past, what makes this study unique is that it reveals a strong \ion{H}{i} absorber for which no corresponding galaxy has been found, despite being sufficiently chemically enriched as indicated by associated metal lines with super-solar metallicity.
Although our initial goal is to map the large-scale \ion{H}{I} distribution, the discovery of this extreme absorber provides a unique window into the violent baryon cycle of a forming galaxy cluster. To definitively resolve the exact nature of this system, further observational efforts are required. 
First, higher-resolution optical spectroscopy of the background quasar is necessary to accurately decompose the complex absorption features. Moreover, deep observations of the Ly$\beta$ region would be beneficial.
Second, deep near-infrared spectroscopy, potentially targeting [\ion{O}{III}] emitters or H$\alpha$ emitters, will be essential to identify the elusive host galaxies missed by our rest-UV selection. 
Furthermore, spatially high-resolution sub-millimetre/radio observations could unveil the detailed kinematics of the possible dust-obscured components. 

\section*{Acknowledgements}
We are grateful to Kazuhiro Shimasaku, Nao Suzuki, Khee-Gan Lee, Roderik Overzier, and Yongming Liang for their insightful comments and helpful suggestions. 
NK was supported by the Japan Society for the Promotion of Science (JSPS) KAKENHI grant (25H00663, 25K01038, 25K01044). 
RS was also supported by the JSPS KAKENHI grant number JP25K01044.
SS is supported by the JSPS KAKENHI grant number JP26KJ0916. 
KI acknowledges support from the Independent Research Fund Denmark (DFF) under grant 3120-00043B. The Cosmic Dawn Center (DAWN) is funded by the Danish National Research Foundation under grant No. 140. 

Data analysis was in part carried out on the Multi-wavelength Data Analysis System operated by the Astronomy Data Center (ADC), National Astronomical Observatory of Japan. 

The Hyper Suprime-Cam (HSC) collaboration includes the astronomical communities of Japan and Taiwan, and Princeton University. The HSC instrumentation and software were developed by the National Astronomical Observatory of Japan (NAOJ), the Kavli Institute for the Physics and Mathematics of the Universe (Kavli IPMU), the University of Tokyo, the High Energy Accelerator Research Organization (KEK), the Academia Sinica Institute for Astronomy and Astrophysics in Taiwan (ASIAA), and Princeton University. Funding was contributed by the FIRST program from Japanese Cabinet Office, the Ministry of Education, Culture, Sports, Science and Technology (MEXT), the Japan Society for the Promotion of Science (JSPS), Japan Science and Technology Agency (JST), the Toray Science Foundation, NAOJ, Kavli IPMU, KEK, ASIAA, and Princeton University. 
This paper makes use of software developed for the Large Synoptic Survey Telescope. We thank the LSST Project for making their code available as free software at \url{http://dm.lsst.org}.
The Pan-STARRS1 Surveys (PS1) have been made possible through contributions of the Institute for Astronomy, the University of Hawaii, the Pan-STARRS Project Office, the Max-Planck Society and its participating institutes, the Max Planck Institute for Astronomy, Heidelberg and the Max Planck Institute for Extraterrestrial Physics, Garching, The Johns Hopkins University, Durham University, the University of Edinburgh, Queen's University Belfast, the Harvard-Smithsonian Center for Astrophysics, the Las Cumbres Observatory Global Telescope Network Incorporated, the National Central University of Taiwan, the Space Telescope Science Institute, the National Aeronautics and Space Administration under Grant No. NNX08AR22G issued through the Planetary Science Division of the NASA Science Mission Directorate, the National Science Foundation under Grant No. AST-1238877, the University of Maryland, and Eotvos Lorand University (ELTE) and the Los Alamos National Laboratory.
Based in part on data collected at the Subaru Telescope and retrieved from the HSC data archive system, which is operated by Subaru Telescope and Astronomy Data Center at National Astronomical Observatory of Japan. 

These data were obtained and processed as part of the CFHT Large Area $U$-band Deep Survey (CLAUDS), which is a collaboration between astronomers from Canada, France, and China described in \citet{Sawicki+19}. CLAUDS is based on observations obtained with MegaPrime/MegaCam, a joint project of CFHT and CEA/DAPNIA, at the CFHT which is operated by the National Research Council
(NRC) of Canada, the Institut National des Science de l'Univers of the Centre National de la Recherche Scientifique (CNRS) of France, and the University of Hawaii. CLAUDS uses data obtained in part through the Telescope Access Program (TAP), which has been funded by the National Astronomical Observatories, Chinese Academy of Sciences, and the Special Fund for Astronomy from the Ministry of Finance of China. CLAUDS uses data products from TERAPIX and the Canadian Astronomy Data Centre (CADC) and was carried
out using resources from Compute Canada and Canadian Advanced Network For Astrophysical Research (CANFAR).

%%%%%%%%%%%%%%%%%%%%%%%%%%%%%%%%%%%%%%%%%%%%%%%%%%
\section*{Data Availability}
The data underlying this article will be shared on reasonable request to the corresponding author.



\bibliographystyle{mnras}
\bibliography{reference} % if your bibtex file is called example.bib

%%%%%%%%%%%%%%%%%%%%%%%%%%%%%%%%%%%%%%%%%%%%%%%%%%

%%%%%%%%%%%%%%%%% APPENDICES %%%%%%%%%%%%%%%%%%%%%

\appendix
\section{All Spectra}
\begin{figure}
    \centering
    \includegraphics[width=\columnwidth]{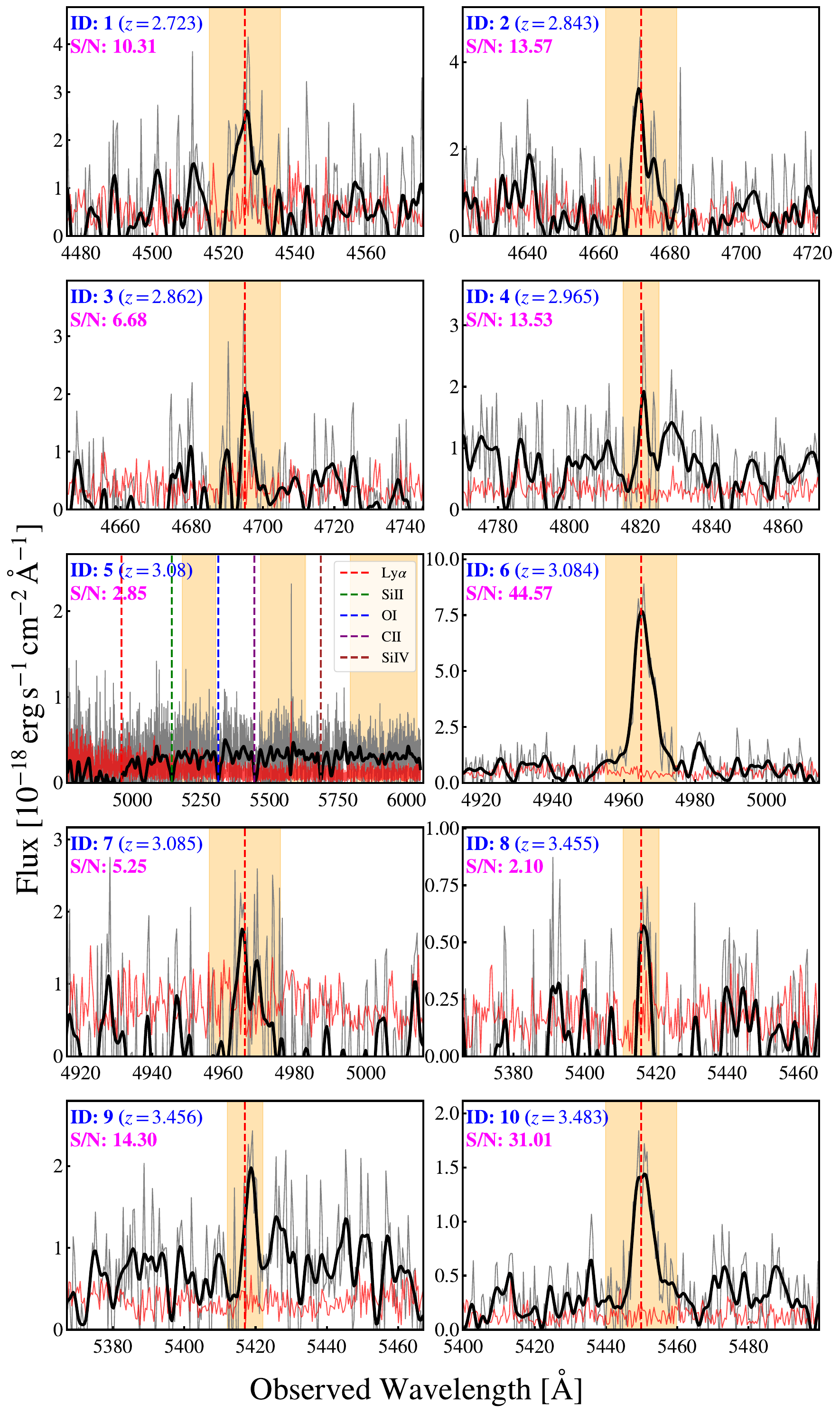}
    \caption{1D spectra of the sources not included in Toshikawa et al.\ (2026, in prep.). The observed data, error spectra, and Gaussian-smoothed spectra ($\mathrm{FWHM} = 2 \, \mathrm{pixels}$, except for ID~5 where $\mathrm{FWHM} = 10 \, \mathrm{pixels}$ for visualization purposes) are shown as grey, red, and black lines, respectively. The S/N, evaluated within the shaded region, is indicated in each panel. Notably, because ID~5 lacks Ly$\alpha$ emission and shows only a Lyman break, its S/N is determined from the continuum level.}
    \label{fig:all_spectra}
\end{figure}

We show all sources spectra not presented in Toshikawa et al. (2026, in prep.) in Figure \ref{fig:all_spectra}.

\section{Other fitting results for multi-component model}
\label{app:other_component_results}
\begin{figure}
    \centering
    \includegraphics[width=\columnwidth]{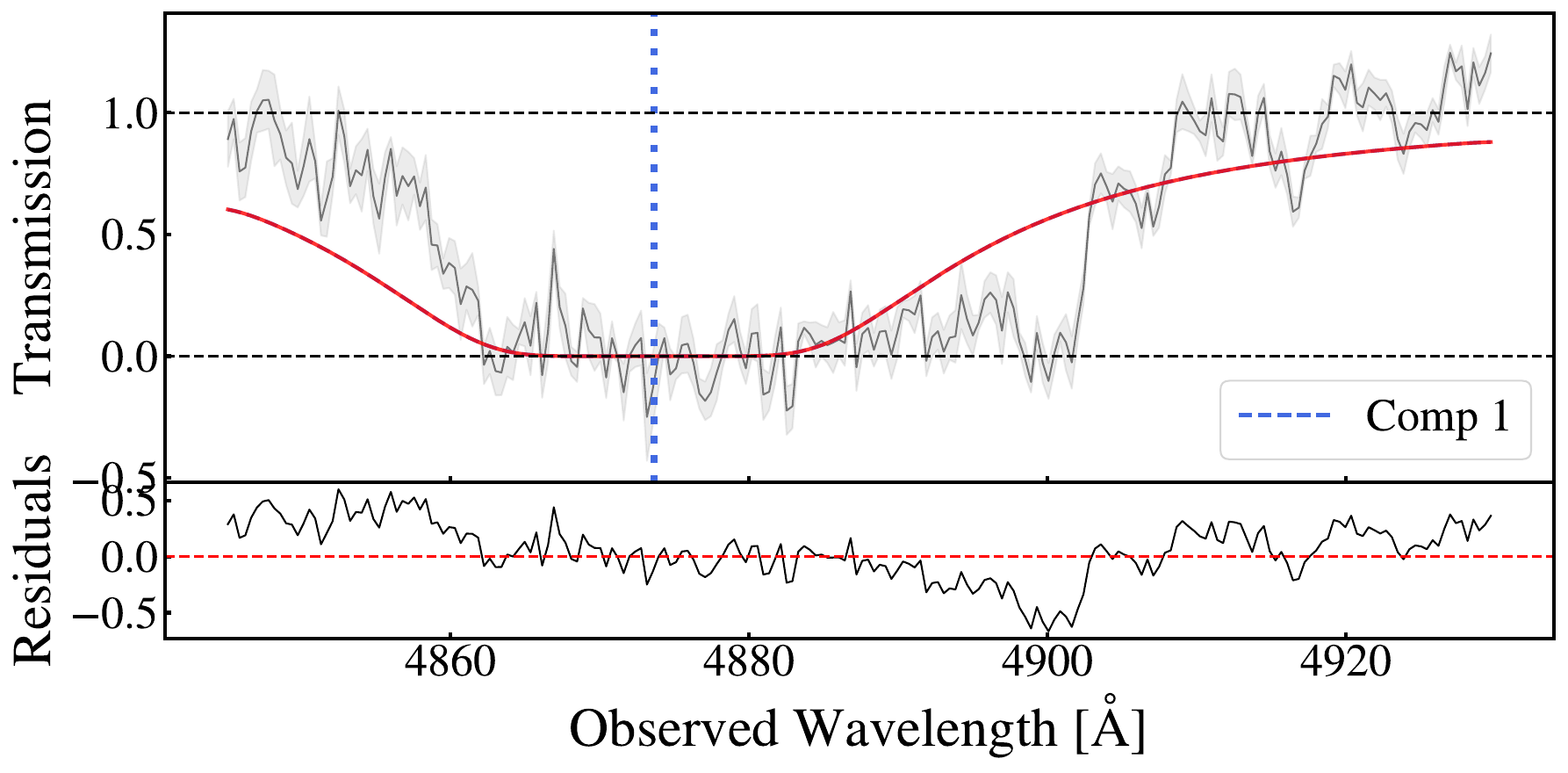}
    \includegraphics[width=\columnwidth]{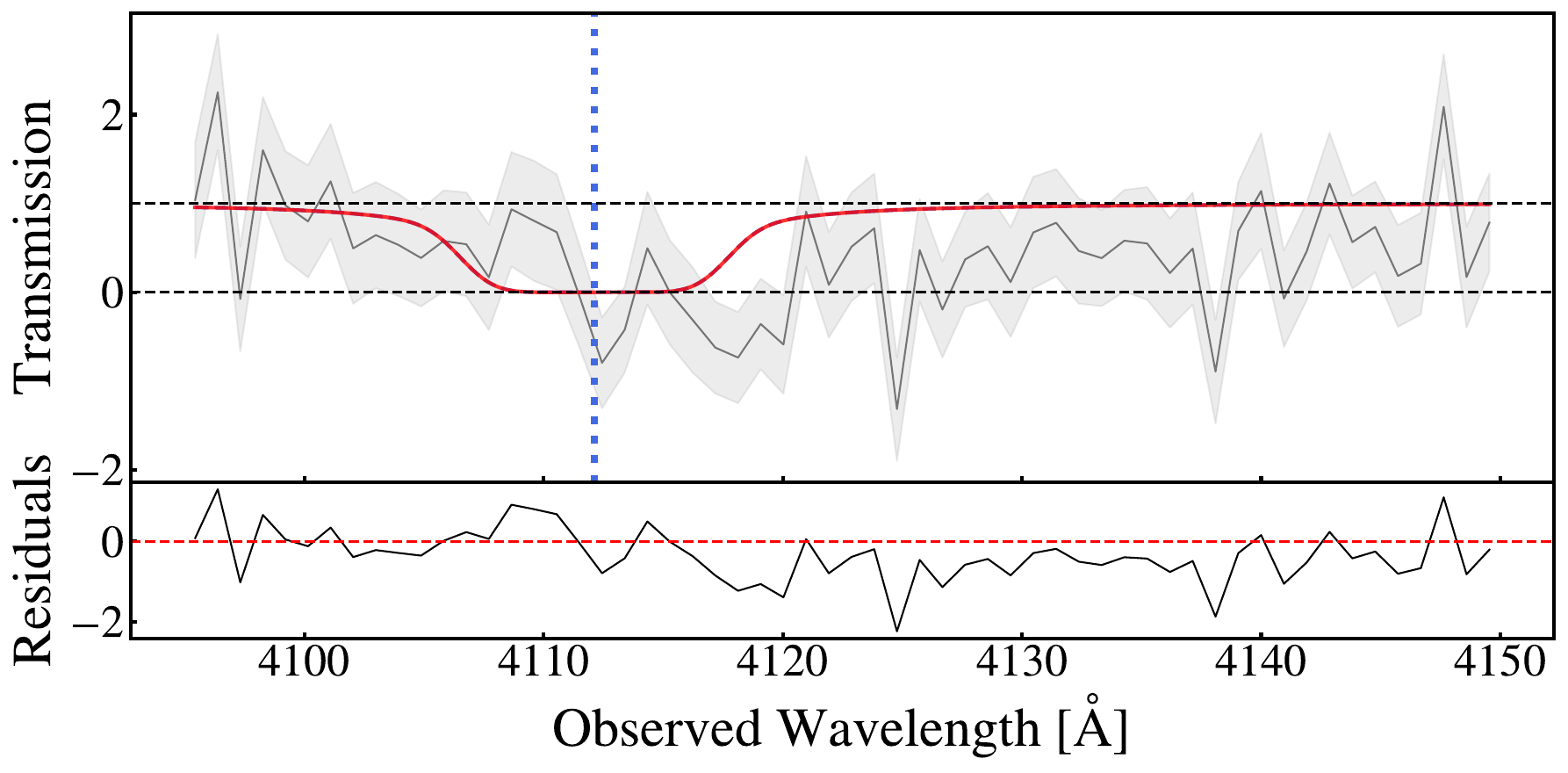}
    \caption{Best-fit one-component Voigt profile model.}
    \label{fig:1component_fit_picture}
\end{figure}

\begin{figure}
    \centering
    \includegraphics[width=\columnwidth]{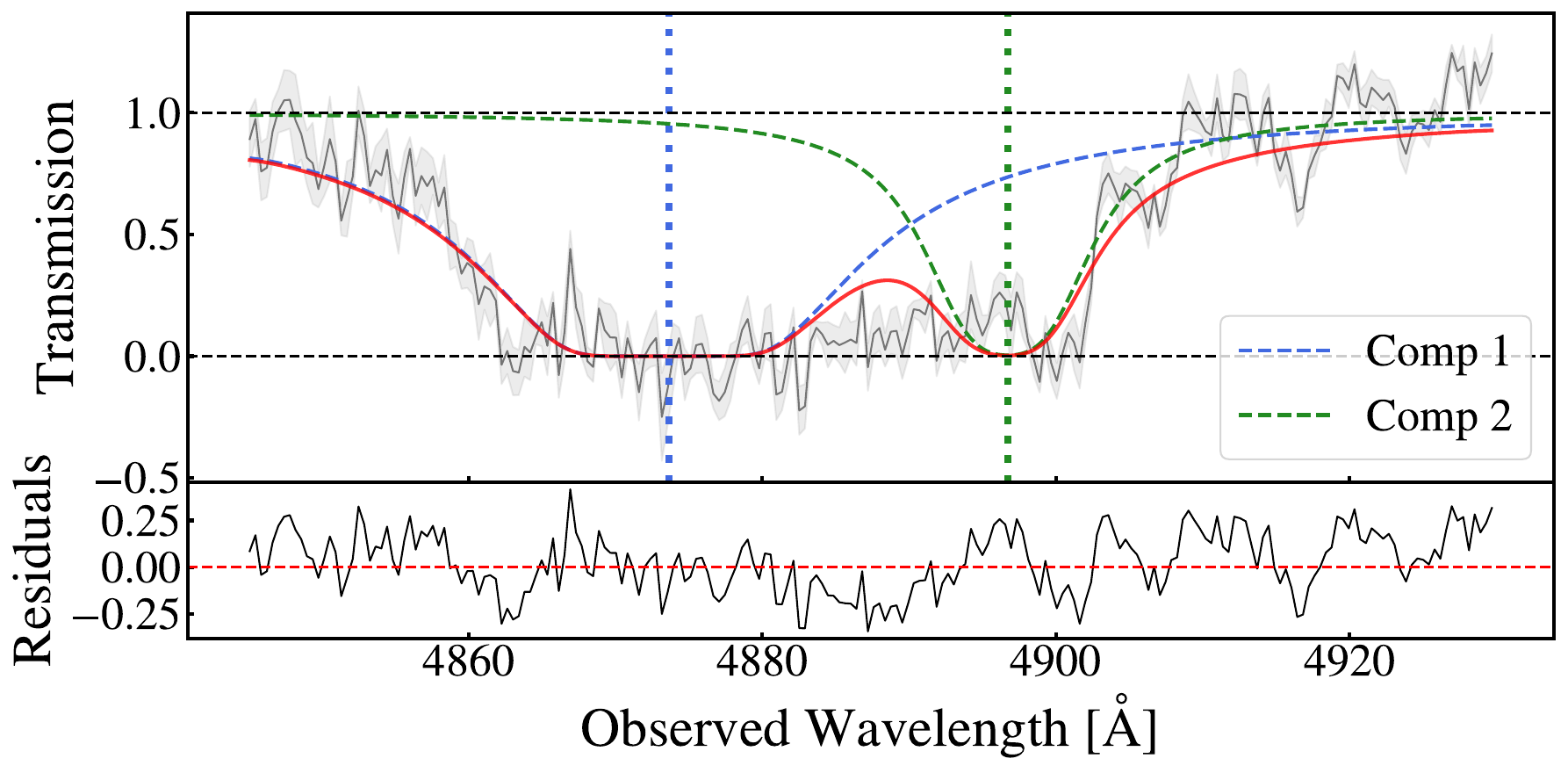}
    \includegraphics[width=\columnwidth]{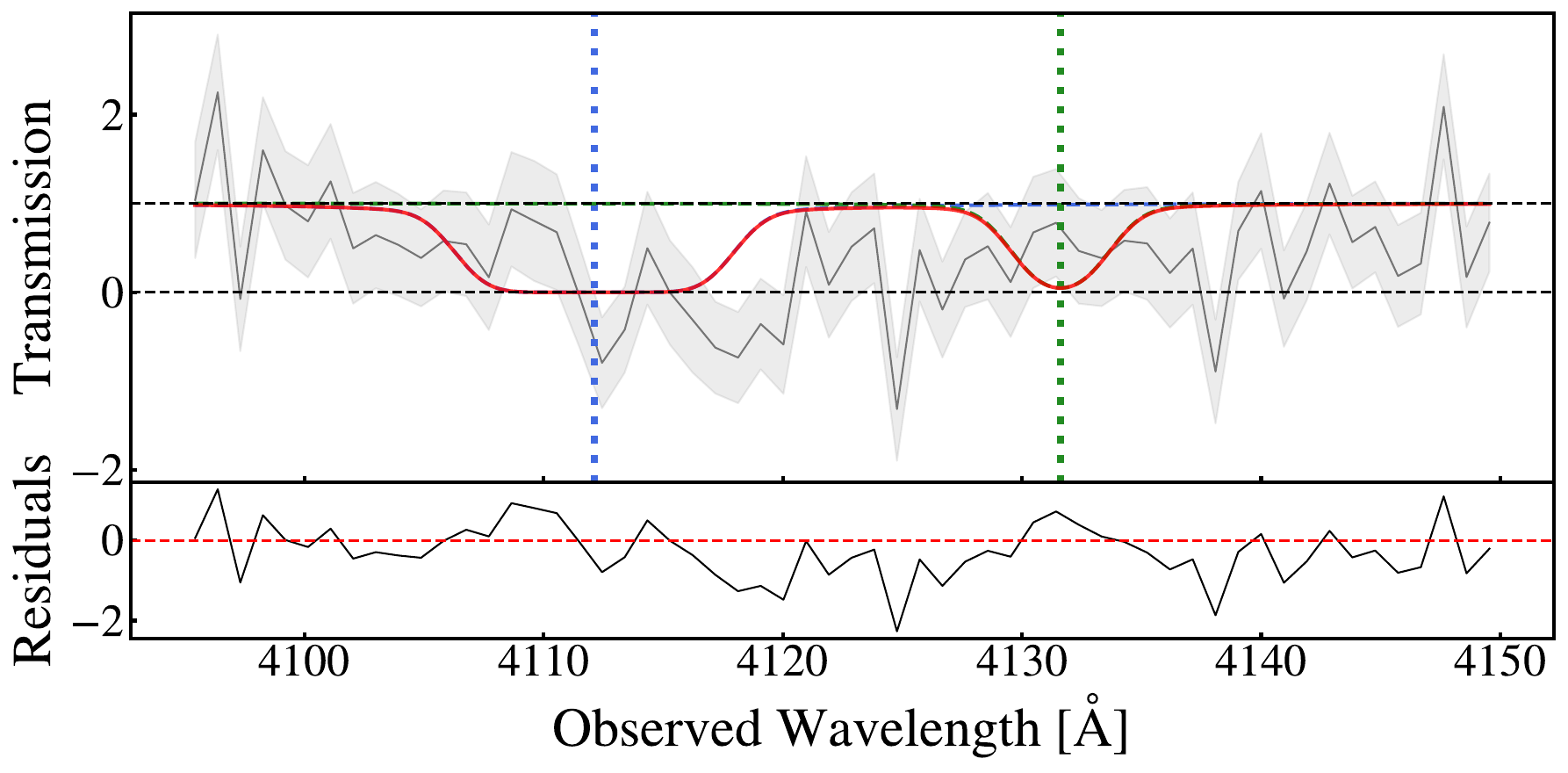}
    \caption{Best-fit two-component Voigt profile model.}
    \label{fig:2component_fit_picture}
\end{figure}

\begin{figure}
    \includegraphics[width=\columnwidth]{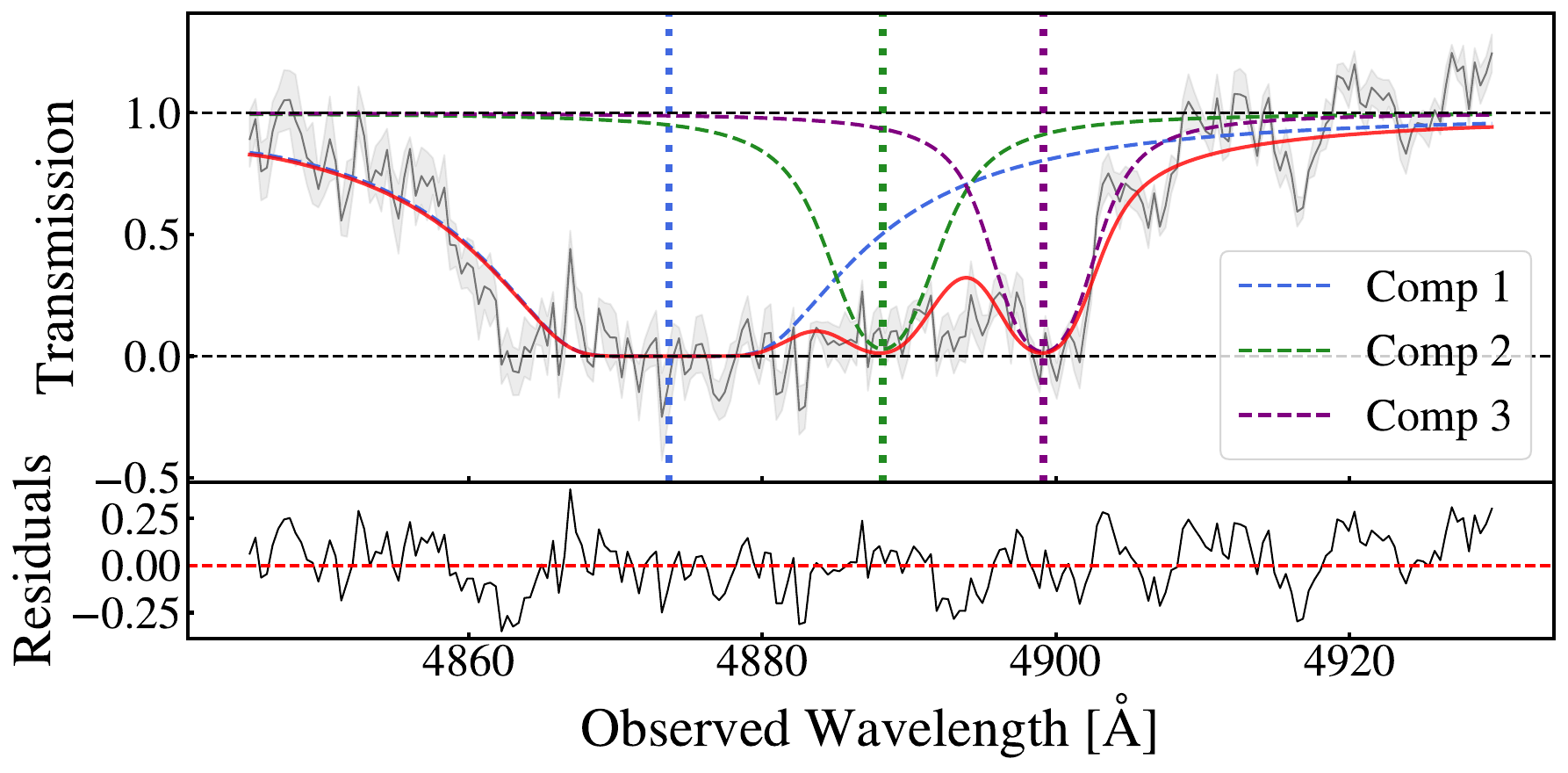}
    \includegraphics[width=\columnwidth]{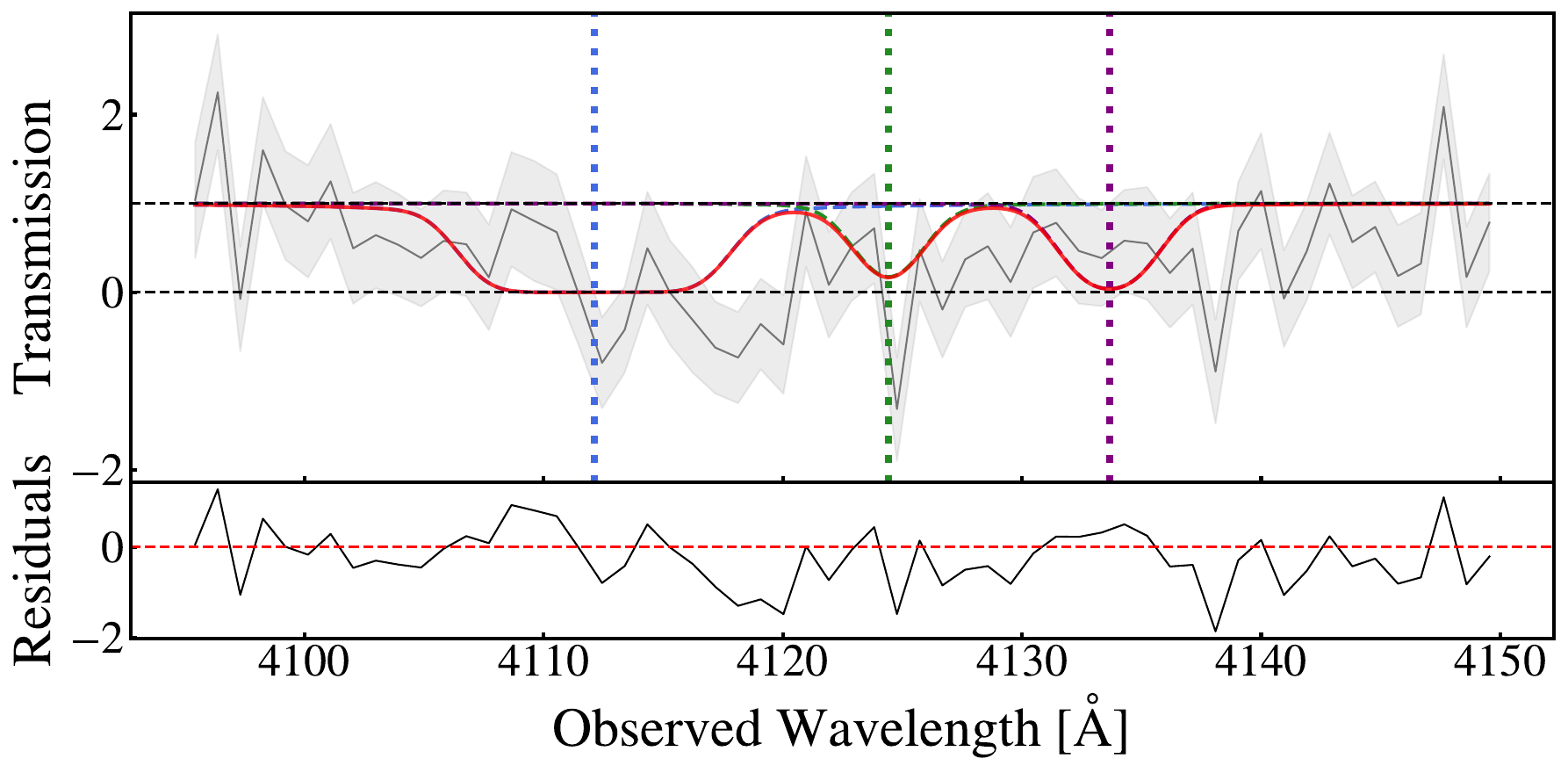}
    \caption{Best-fit three-component Voigt profile model.}
    \label{fig:3component_fit_picture}
\end{figure}

\begin{figure}
    \centering
    \includegraphics[width=\columnwidth]{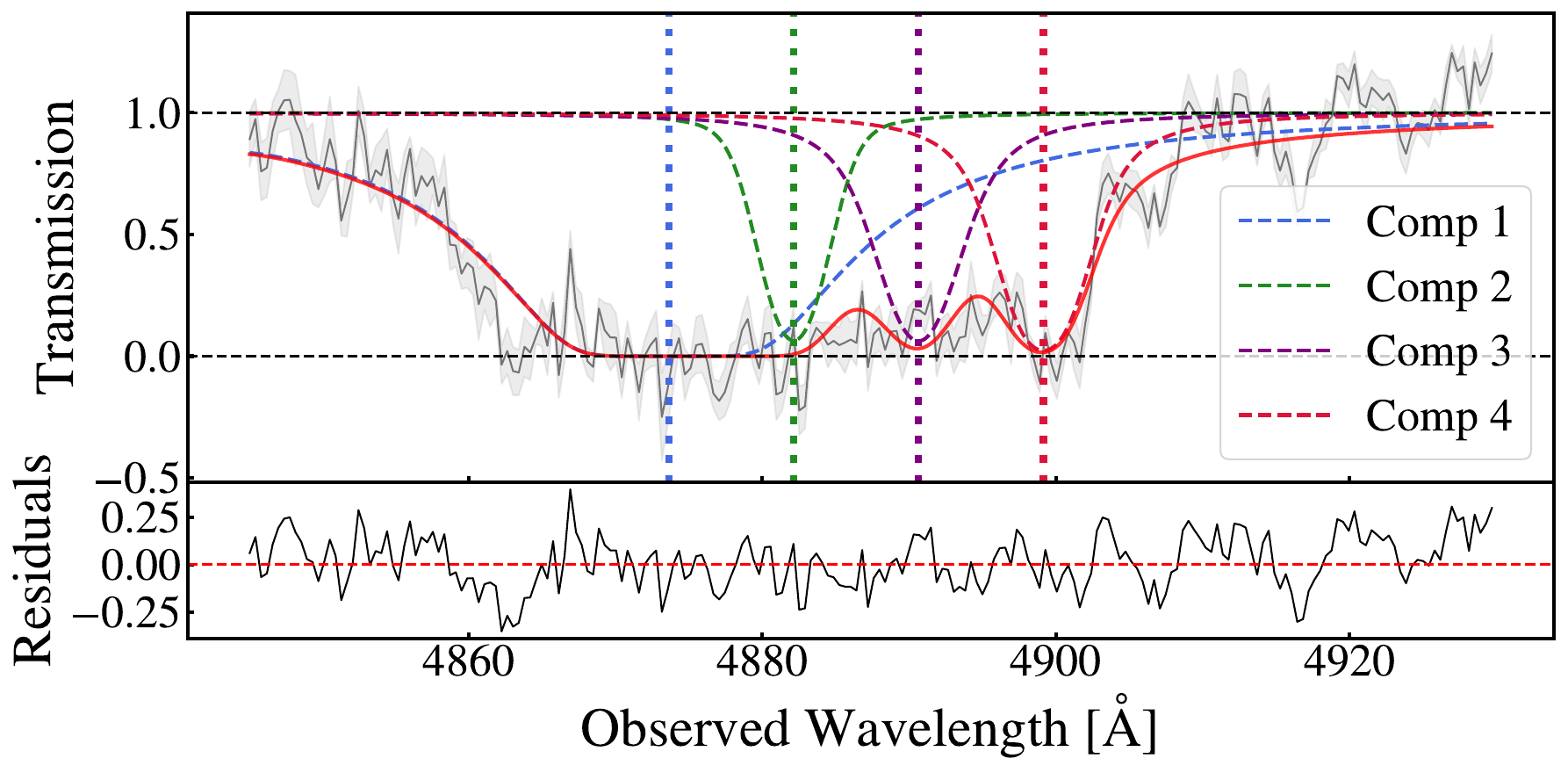}
    \includegraphics[width=\columnwidth]{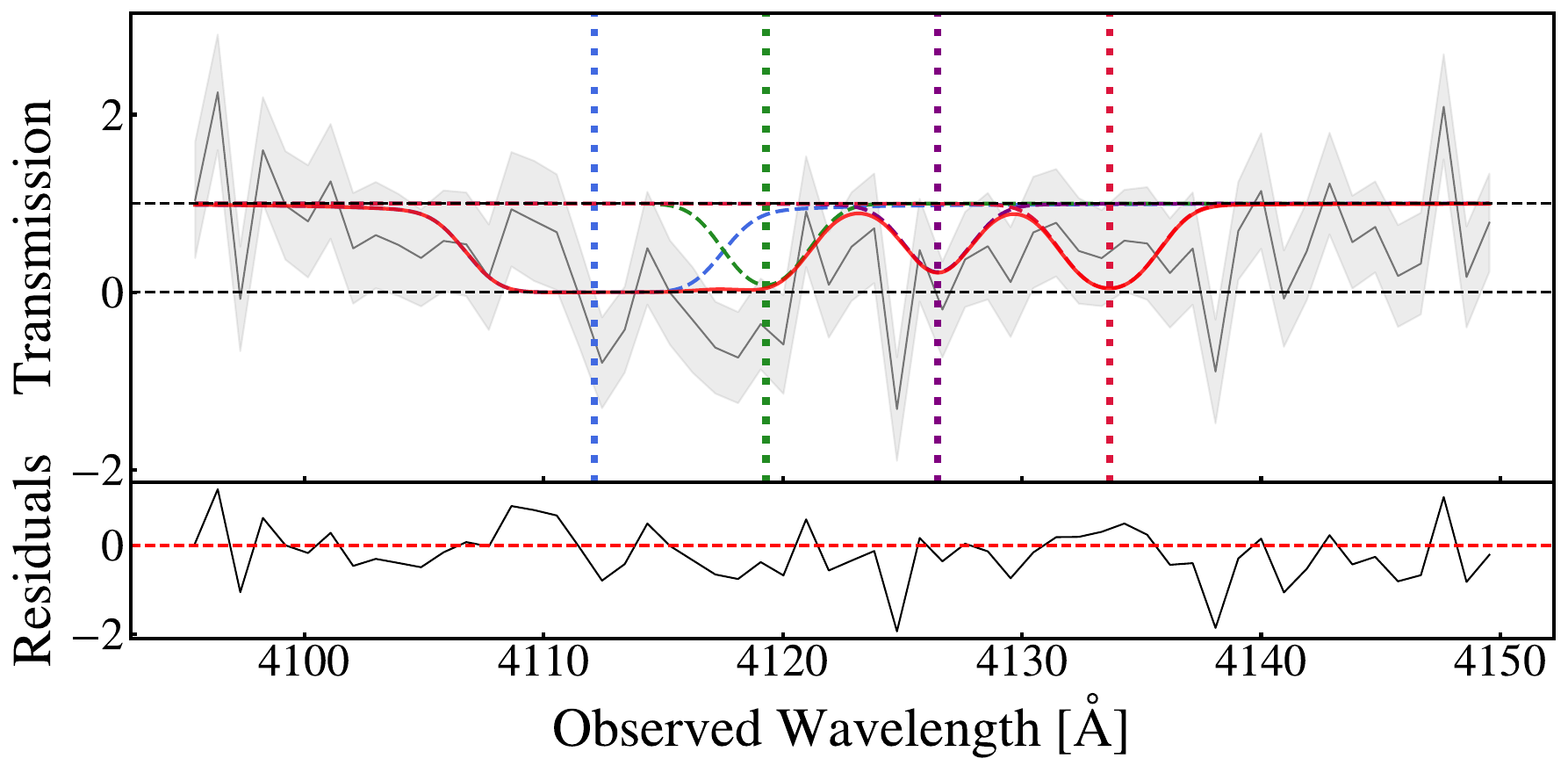}
    \caption{Best-fit four-component Voigt profile model.}
    \label{fig:4component_fit_picture}
\end{figure}

\begin{figure}
    \centering
    \includegraphics[width=\columnwidth]{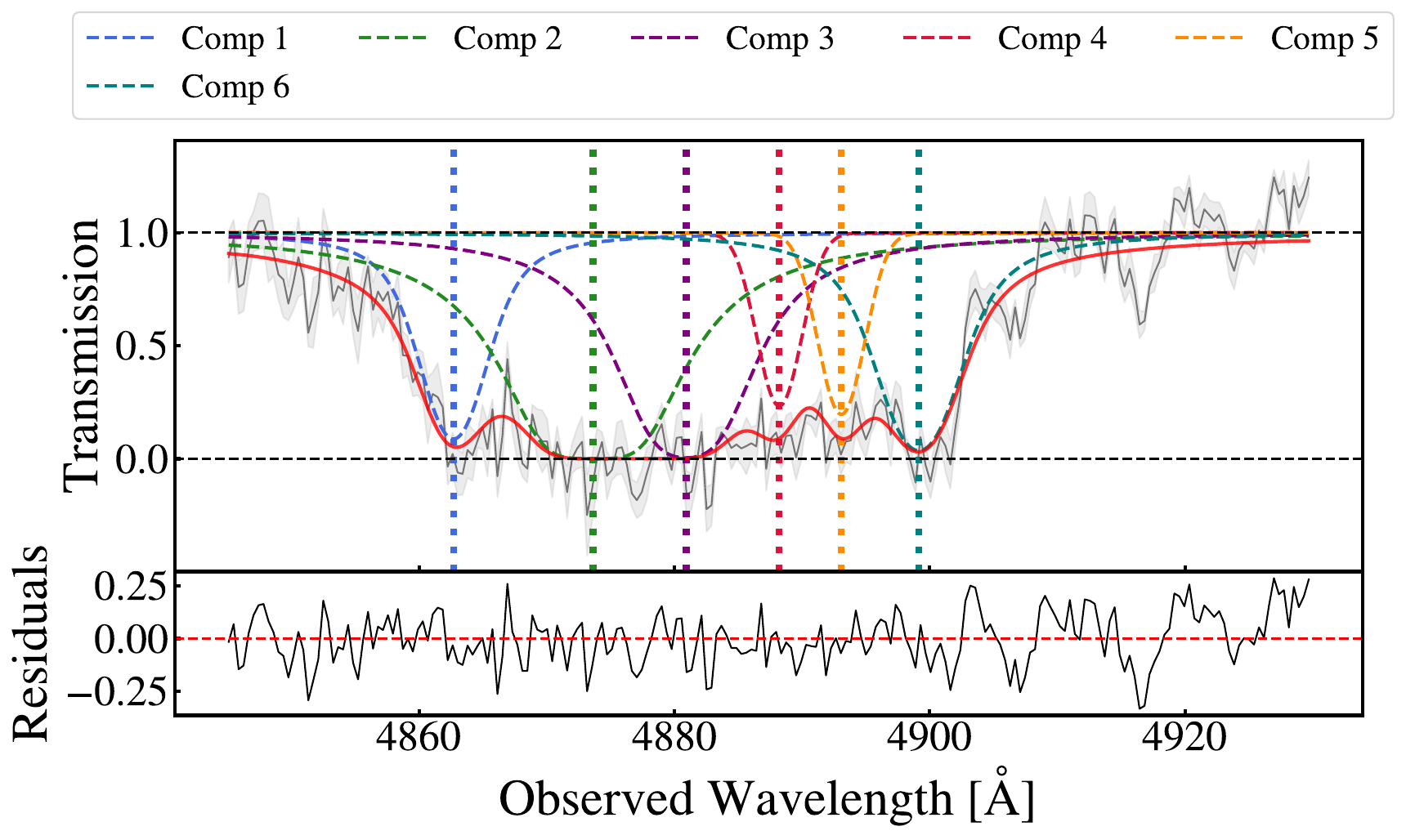}
    \includegraphics[width=\columnwidth]{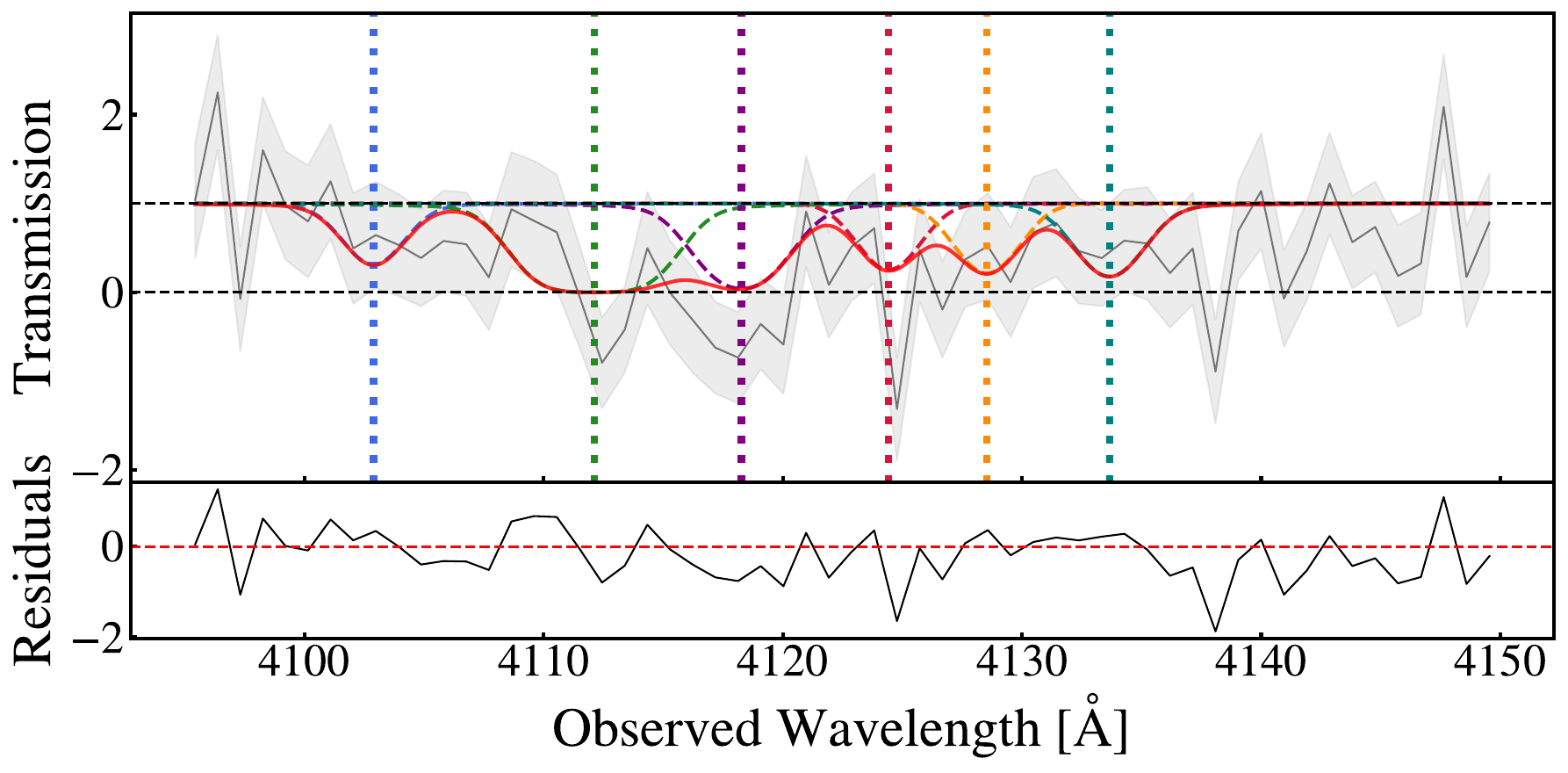}
    \caption{Best-fit six-component Voigt profile model.}
    \label{fig:6component_fit_picture}
\end{figure}

\begin{figure}
    \centering
    \includegraphics[width=\columnwidth]{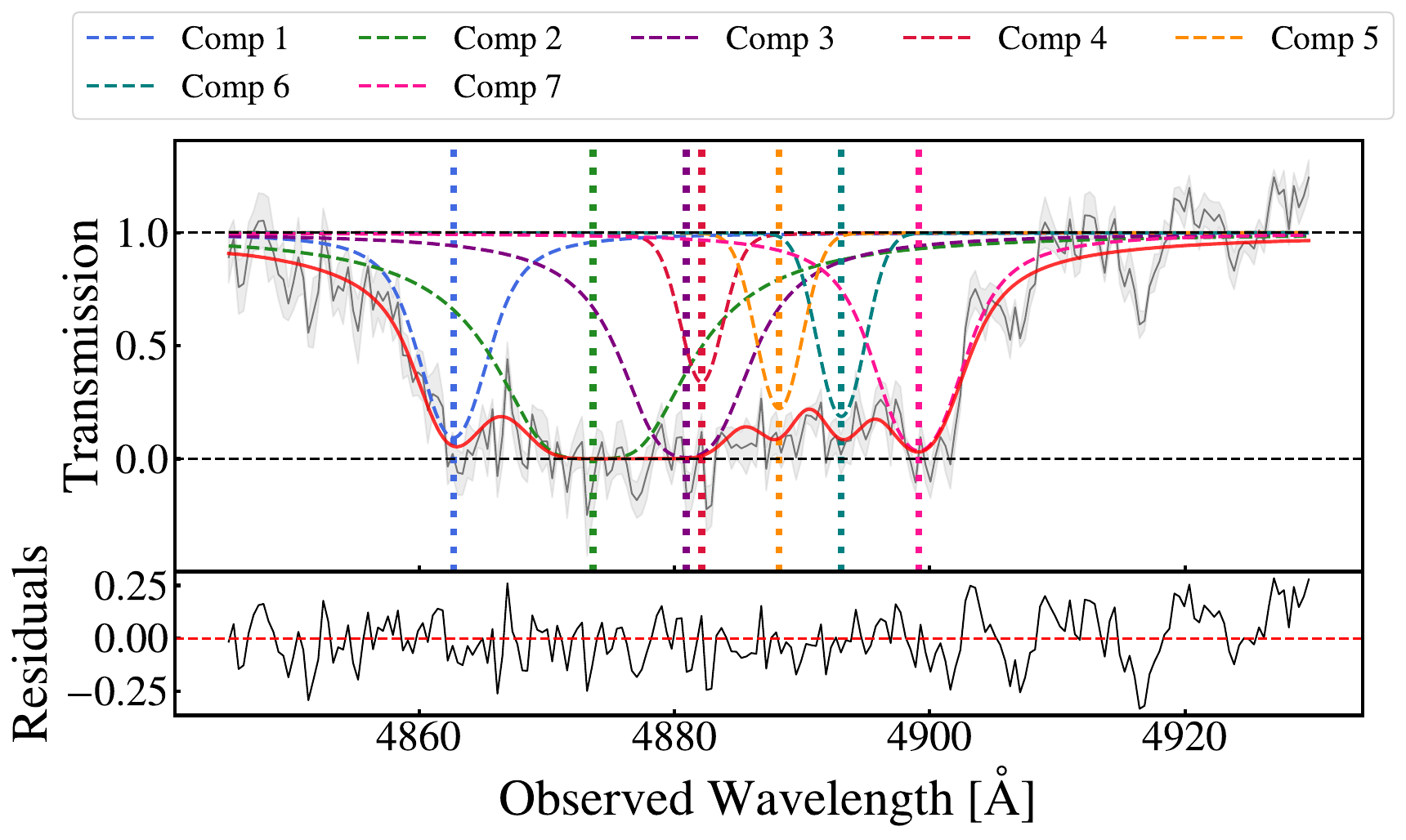}
    \includegraphics[width=\columnwidth]{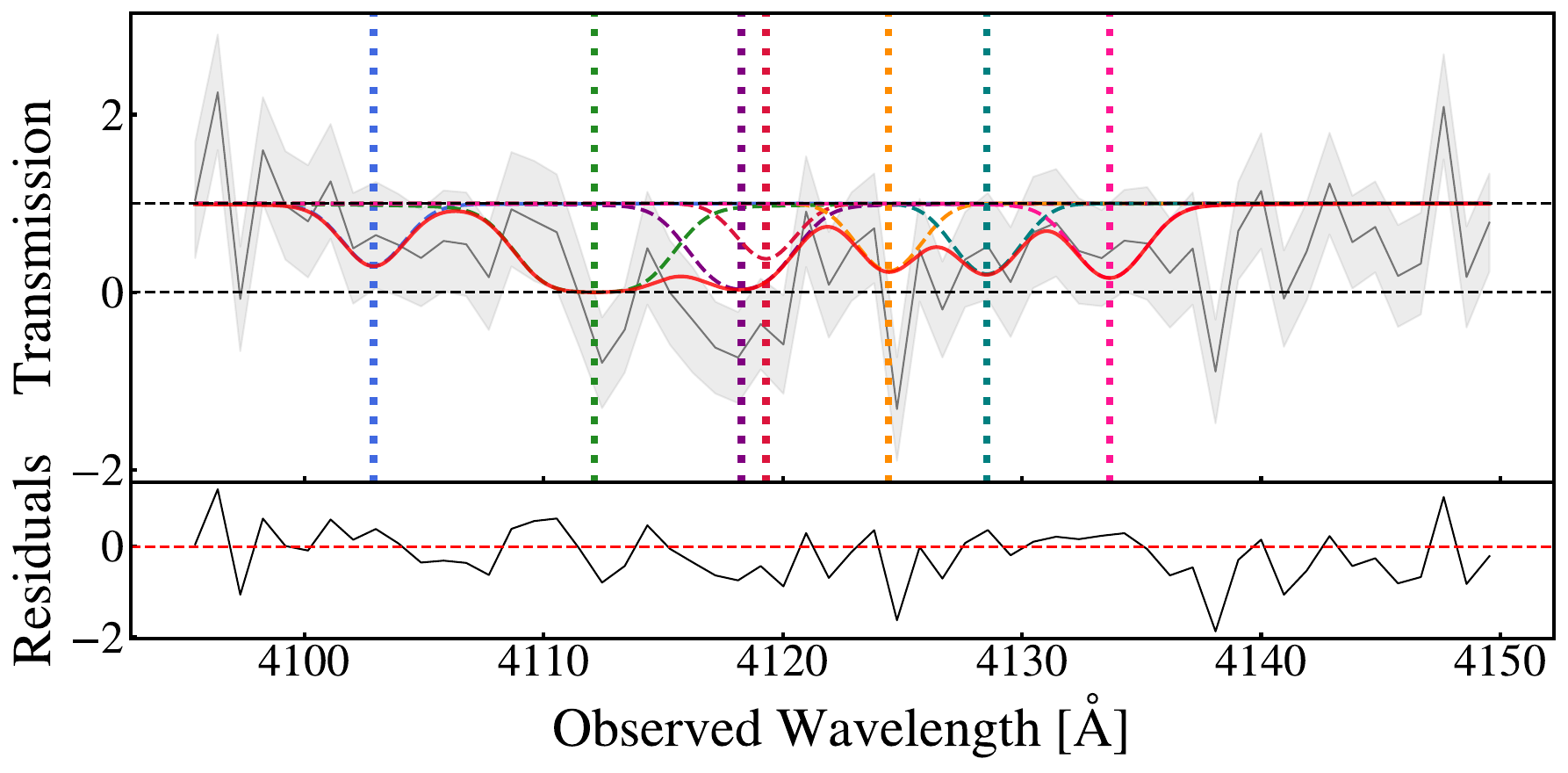}
    \caption{Best-fit seven-component Voigt profile model.}
    \label{fig:7component_fit_picture}
\end{figure}

We show the result of the one-four, six and seven-component fit for the strongest \ion{H}{I} absorption in Figure \ref{fig:1component_fit_picture}, \ref{fig:2component_fit_picture}, \ref{fig:3component_fit_picture}, \ref{fig:4component_fit_picture}, \ref{fig:6component_fit_picture} and \ref{fig:7component_fit_picture}. The fitted parameters for each model are shown in Table \ref{tab:other_component_results}.

\begin{table}
    \centering
    \caption{MCMC fitting results for the 1, 2, 3, 4, 6, and 7-component models.}
    \label{tab:other_component_results}
    \renewcommand{\arraystretch}{1.4}
    \begin{tabular}{c c c c} \hline \hline
        Number  & $z$ & $\log N_{\mathrm{H\,I}}$ [cm$^{-2}$] & $b$ [km s$^{-1}$] \\ \hline
        1       & 3.009 (fixed)                   & $20.76^{+0.01}_{-0.01}$ & $107.2^{+10.4}_{-47.2}$ \\ \hline
        2       & 3.009 (fixed)                   & $20.37^{+0.02}_{-0.01}$ & $117.7^{+2.3}_{-5.9}$   \\ 
                & $3.028^{+0.0001}_{-0.0002}$ & $19.55^{+0.01}_{-0.03}$ & $42.0^{+6.4}_{-23.2}$   \\ \hline
        3       & 3.009 (fixed)                   & $20.31^{+0.02}_{-0.02}$ & $116.0^{+3.1}_{-6.4}$   \\
                & $3.021^{+0.0004}_{-0.0004}$ & $19.19^{+0.09}_{-0.10}$ & $29.3^{+8.9}_{-9.7}$    \\
                & $3.030^{+0.0002}_{-0.0002}$ & $19.05^{+0.07}_{-0.08}$ & $47.8^{+1.6}_{-3.9}$    \\ \hline
        4       & 3.009 (fixed)                   & $20.31^{+0.02}_{-0.02}$ & $108.4^{+8.0}_{-18.5}$  \\
                & $3.016^{+0.0005}_{-0.0005}$ & $18.42^{+0.55}_{-1.01}$ & $43.6^{+4.7}_{-8.8}$    \\
                & $3.023^{+0.0002}_{-0.0007}$ & $18.97^{+0.16}_{-0.19}$ & $26.7^{+10.6}_{-8.4}$   \\
                & $3.030^{+0.0002}_{-0.0002}$ & $18.98^{+0.11}_{-0.13}$ & $46.7^{+2.4}_{-5.0}$    \\ \hline
        6       & $3.000^{+0.0002}_{-0.0002}$ & $18.88^{+0.12}_{-0.15}$ & $22.1^{+12.4}_{-8.3}$   \\
                & 3.009 (fixed)                   & $19.81^{+0.10}_{-0.18}$ & $75.1^{+18.3}_{-27.0}$  \\
                & $3.015^{+0.0006}_{-0.0006}$ & $19.54^{+0.19}_{-0.22}$ & $44.1^{+4.3}_{-9.5}$    \\
                & $3.021^{+0.0002}_{-0.0003}$ & $17.02^{+1.05}_{-1.03}$ & $34.8^{+7.8}_{-8.3}$    \\
                & $3.025^{+0.0002}_{-0.0002}$ & $17.05^{+1.02}_{-0.94}$ & $37.8^{+7.6}_{-7.9}$    \\
                & $3.030^{+0.0002}_{-0.0002}$ & $19.13^{+0.06}_{-0.07}$ & $29.3^{+10.6}_{-12.5}$  \\ \hline
        7       & $3.000^{+0.0002}_{-0.0002}$ & $18.85^{+0.12}_{-0.17}$ & $23.0^{+12.0}_{-8.0}$   \\
                & 3.009 (fixed)                   & $19.84^{+0.10}_{-0.14}$ & $69.4^{+17.7}_{-25.3}$  \\
                & $3.015^{+0.0008}_{-0.0008}$ & $19.45^{+0.26}_{-0.43}$ & $47.1^{+2.3}_{-7.7}$    \\
                & $3.016^{+0.0008}_{-0.0012}$ & $17.62^{+1.09}_{-1.46}$ & $23.7^{+16.3}_{-8.9}$   \\
                & $3.021^{+0.0003}_{-0.0003}$ & $17.23^{+0.92}_{-1.00}$ & $34.5^{+7.8}_{-7.7}$    \\
                & $3.025^{+0.0002}_{-0.0003}$ & $17.11^{+1.00}_{-1.02}$ & $38.3^{+6.9}_{-8.3}$    \\
                & $3.030^{+0.0002}_{-0.0002}$ & $19.13^{+0.07}_{-0.07}$ & $30.6^{+8.8}_{-13.8}$   \\ \hline
    \end{tabular}
\end{table}

\section{The corner plot for five-component model fitting}
\label{app:corner_plot}
\begin{figure*}
    \centering
    \includegraphics[width=\textwidth]{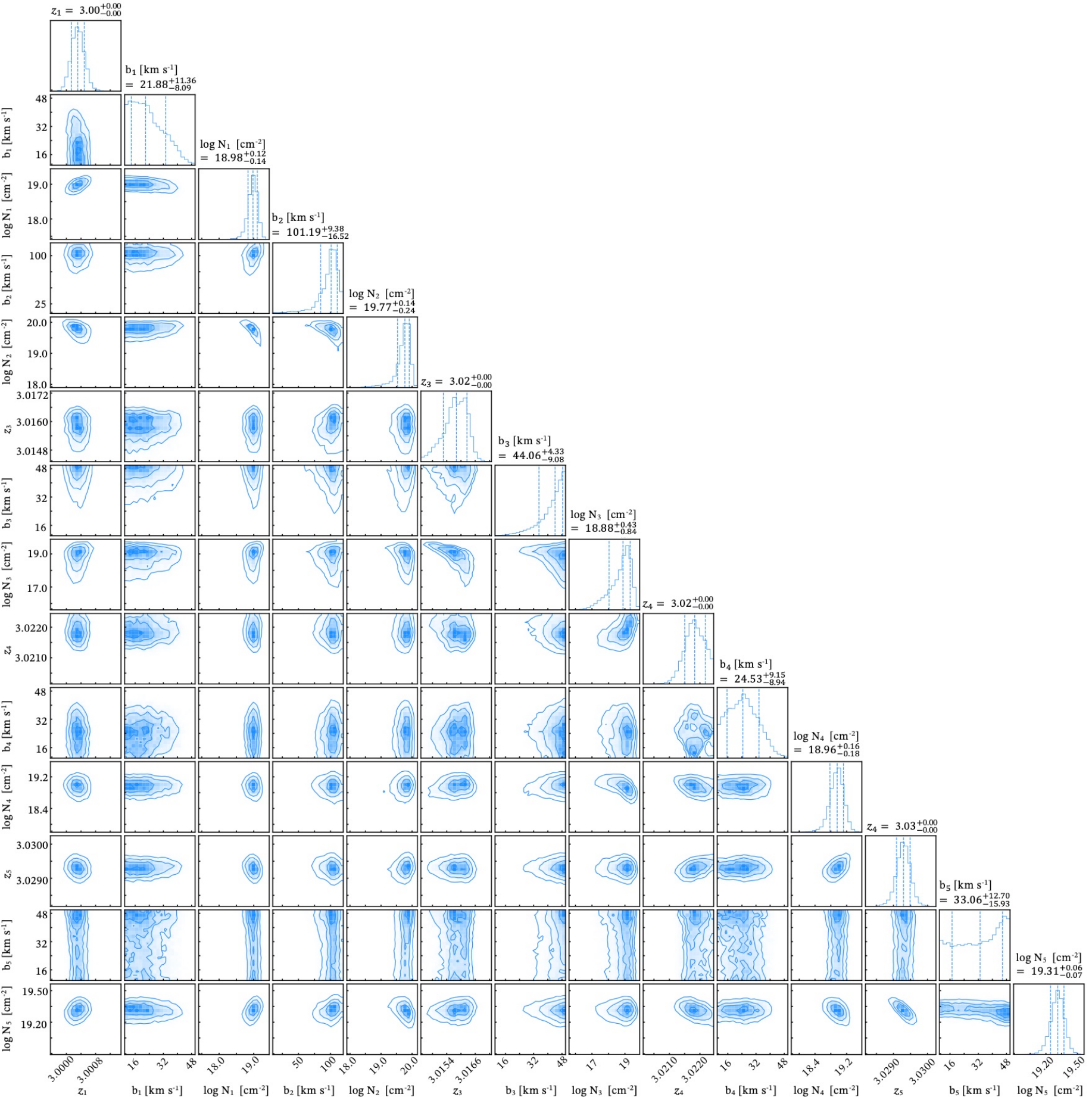}
    \caption{Corner plot of the MCMC posterior probability distributions for the five-component Voigt profile fitting. $(z_n, N_n, b_n)$ represent the best-fit parameters for Component $n$. The 1D histograms along the diagonal display the marginalized distribution for each parameter, with vertical dashed lines indicating the 16th, 50th (median), and 84th percentiles. The off-diagonal 2D contour plots illustrate the covariances between parameter pairs.}
    \label{fig:corner_plot_5}
\end{figure*}

The full posterior probability distributions for the five-component model fit are shown in Figure \ref{fig:corner_plot_5}.

%%%%%%%%%%%%%%%%%%%%%%%%%%%%%%%%%%%%%%%%%%%%%%%%%%

% Don't change these lines
\bsp	% typesetting comment
\label{lastpage}
\end{document}